\documentclass[a4paper,11pt]{article}
\usepackage{jcappub} 
\usepackage{graphics}
\usepackage{comment}
\usepackage{subcaption}
\usepackage{url}
\usepackage[table,xcdraw,dvipsnames]{xcolor}
\usepackage[normalem]{ulem}
\usepackage{float}
\usepackage{hyperref}
\usepackage{orcidlink} 
\usepackage[normalem]{ulem} 

\usepackage{bbm}
\definecolor{blue-violet}{rgb}{0.7, 0.2, 0.8}

\makeatletter
\gdef\@fpheader{}
\makeatother

\title{Coherent versus incoherent amplification: from parametric resonance to the kinetic regime in stimulated axion decay}

\newcommand{\dfa}{Dipartimento di Fisica e Astronomia ‘G. Galilei’, Università di Padova, Via Francesco Marzolo 8, 35131 Padova, Italy}
\newcommand{\infnpd}{INFN Sezione di Padova, Via Francesco Marzolo 8, 35131 Padova, Italy}

\author[1,2]{Clemente Smarra~\orcidlink{0000-0002-0817-2830}}
\emailAdd{clemente.smarra@unipd.it}
\affiliation[1]{\dfa}
\affiliation[2]{\infnpd}

\author[3,4,5]{Elisa Todarello~\orcidlink{0000-0001-7419-0976}}
\emailAdd{elisa.todarello@berkeley.edu}
\affiliation[3]{Leinweber Institute for Theoretical Physics, University of California, Berkeley, CA 94720, U.S.A.}
\affiliation[4]{Theoretical Physics Group, Lawrence Berkeley National Laboratory, Berkeley, CA 94720, U.S.A.}
\affiliation[5]{Universit\`a degli Studi di Torino, via P. Giuria 1, 10125 Torino, Italy}

\abstract{Although the spontaneous decay of axions into two photons is extremely slow on cosmological timescales, it can be strongly enhanced in environments with large bosonic occupation numbers. From a classical field perspective, stimulated axion decay is described by modified Maxwell equations. For a coherent axion background, these equations exhibit a parametric resonance characterized by a photon growth exponent $s\propto g_{a\gamma}$. By contrast, a standard kinetic description, for instance based on Boltzmann equations, involves the squared matrix element of the decay process and therefore would naturally predict a characteristic rate scaling as $s\propto g_{a\gamma}^2$. In this work, we show that these apparently different behaviors originate from the coherence properties of the axion field. Starting from an axion field with a finite momentum dispersion, we show how dephasing among its momentum modes modifies the photon instability and derive an effective evolution equation in which this effect is encoded in a kernel. 
For a continuous Maxwell-Boltzmann distribution of axion velocities, we obtain an implicit expression for the photon growth exponent that reproduces the two asymptotic regimes and provides an approximate description of the crossover.
When amplification develops on scales shorter than the axion coherence length, the parametric resonance scaling $s\propto g_{a\gamma}$ is recovered, whereas in the opposite limit an effectively kinetic behavior with $s\propto g_{a\gamma}^2$ emerges. Although the detailed form of the crossover depends on the axion momentum distribution, we expect the asymptotic scaling with $g_{a\gamma}$ to be largely insensitive to the specific distribution assumed.}

\begin{document}

\maketitle
\flushbottom

\section{Introduction}
The Standard Model provides an extraordinarily successful description of particle interactions, yet several fundamental questions remain unresolved, including the strong-CP problem. Indeed, although the QCD Lagrangian allows for a CP-violating term, experimental observations constrain CP violation in the strong interactions to be extremely small. The Peccei-Quinn mechanism offers a dynamical explanation for this suppression by introducing an anomalous, spontaneously broken global symmetry whose pseudo-Nambu-Goldstone boson is the axion~\cite{Peccei:1977, Peccei:1977b, Weinberg:1978, Wilczek:1978}. 
Early realizations placed the Peccei-Quinn scale close to the electroweak scale~\cite{Weinberg:1978, Wilczek:1978} and were rapidly excluded by experimental constraints. This motivated the development of ``invisible'' axion models~\cite{Kim:1979, SHIFMAN1980493, DINE1981199, Zhitnitsky:1980tq}, in which the symmetry-breaking scale is much larger and which remain viable over a broad, still largely unexplored region of parameter space; see Ref.~\cite{Di_Luzio_2020} for an excellent review.\\
Despite its light mass, the axion was soon recognized as a compelling candidate for dark matter (DM)~\cite{ABBOTT1983133, PRESKILL1983127, DINE1983137}. Furthermore, string-theory constructions generally predict the presence of additional light pseudoscalar fields with properties similar to those of the QCD axion, but typically unrelated to the strong-CP problem~\cite{Green:1987sp,Svrcek:2006yi,Halverson:2019cmy,Demirtas:2021gsq}. These particles are commonly referred to as axion-like particles (ALPs). In the following, unless otherwise specified, we use the term \textit{axion} to refer collectively to both the QCD axion and ALPs.\\
\newline
A common and phenomenologically important feature of these particles is their coupling to photons, which is modeled by the following Lagrangian density
\begin{equation}
    \mathcal{L}_{a\gamma} = -\frac{1}{4}F_{\mu\nu}F^{\mu\nu} + \frac{1}{2}\partial_\mu a ~\partial^\mu a -\frac{1}{2}m_a^2 a^2 - \frac{1}{4}g_{a\gamma} a F_{\mu\nu}\tilde F^{\mu\nu}.
    \label{eq:lagr}
\end{equation}
Here, $a$ denotes the axion field, $m_a$ and $g_{a\gamma}$ its mass and coupling, respectively, and $F_{\mu\nu} = \partial_\mu A_\nu - \partial_\nu A_\mu$ is the electromagnetic field strength tensor, with $\tilde F^{\mu\nu}$ its dual and $A_\mu$ the vector potential. \\
In the presence of an external electromagnetic field, this interaction allows axions and photons to interconvert, providing one of the main experimental handles for axion searches. For instance, axion-photon conversion is exploited in laboratory experiments aimed at producing and detecting axions~\cite{Anselm:1985obz,VanBibber:1987rq, Ehret:2010mh, Bahre:2013ywa}, in searches for axion DM~\cite{Sikivie:1983ip,Asztalos:2003px,Brun:2019lyf} and in helioscope experiments~\cite{Sikivie:1983ip,vanBibber:1988ge,Armengaud:2019uso}.
The underlying conversion formalism has been studied extensively in a wide range of physical settings; see e.g. Refs.~\cite{Maiani:1986md, Sikivie:1983, Sikivie:1985, 
Raffelt:1987im,
Huang_2018,  
Leroy:2019ghm,
Battye:2019aco, Millar:2021gzs, Carenza:2023nck,McDonald:2023ohd, Smarra:2024vzg}.\\
\newline
The same interaction also allows axions to decay into two photons. For the light axions relevant to DM, the perturbative decay rate is extremely small, $\Gamma_a \propto g_{a\gamma}^2 m_a^3$, implying lifetimes much longer than the age of the Universe and thereby ensuring the stability required of a DM candidate~\cite{ABBOTT1983133, PRESKILL1983127, DINE1983137}. This conclusion can change dramatically in highly occupied bosonic environments. In particular, highly occupied photon states can Bose-enhance the process, leading to stimulated decay~\cite{ABBOTT1983133, PRESKILL1983127}. 
\newline
The possible implications of stimulated axion decay have been investigated in a variety of contexts. In cosmology, the expansion of the Universe and the ambient plasma can significantly affect the dynamics of the  decay~\cite{ABBOTT1983133, PRESKILL1983127, DINE1981199, Lee:2000, Carenza_2020, Alvarez:2020}. A broad range of local astrophysical applications, for which cosmic expansion is irrelevant, has also been considered.  These include radio signals from stimulated axion decay in dense axion environments and axion stars~\cite{Tkachev:1987cd, Hertzberg_2018,Levkov:2020txo, Chen:2023jki, Escudero_2024}, signals from superradiant clouds around rotating black holes~\cite{Rosa_2018,Blas:2020nbs} and axion-echo searches in which an external radio source stimulates the decay of ambient axion DM~\cite{Caputo_2018, Caputo_2019, Ayad:2020fzc, Arza:2019b, Arza:2021nec,Arza:2021zqc,Buen-Abad:2021qvj, Sun:2021oqp,Todarello:2023xuf,Sun:2024, Ghosh:2020hgd, Dev:2023ijb, Guo:2024oqo,Yang:2025yvh}. Related laboratory applications have also been proposed~\cite{An_2026, Liu:2025fvn}.\\
\newline
Throughout this work, we adopt a classical framework and treat both the axion and electromagnetic fields as classical waves.
Within this description, stimulated axion decay manifests as a parametric instability of the electromagnetic field. The axion-photon system is governed by modified Maxwell equations which, for a coherent, zero-momentum axion background, can be mapped onto a Mathieu equation~\cite{Carenza_2020, Alvarez:2020, Hertzberg_2018}. The resulting \textit{parametric resonance} leads to an exponential growth of the  photon field, with a growth exponent that is linear in the axion-photon coupling, namely $s\propto g_{a\gamma}$.\\
The same stimulated decay process can also be approached from a kinetic perspective, for instance through Boltzmann equations for the axion and photon phase-space distributions~\cite{Caputo_2019}. In this language, the collision term governing the evolution of particle occupation numbers is proportional to the squared matrix element of the interaction. Therefore, for the axion decay into two photons, one would naturally be led to expect a characteristic rate scaling as $s\propto g_{a\gamma}^2$ instead. Indeed, in a companion work~\cite{Smarra:2026}, we recover such an exponentially growing solution directly from the Boltzmann equations.\\
\newline
The coexistence of these two different coupling scalings raises a natural question: \textit{when should stimulated axion decay be described by the coherent, parametric-resonance picture, and when should a kinetic description become appropriate?}
We argue that the answer is controlled by the coherence properties of the axion field. If the characteristic lengthscale over which the electromagnetic instability develops is shorter than the axion coherence length, the photon effectively probes a single coherent axion domain. In this regime, the axion can be approximated as a coherent wave, the usual Mathieu description applies, and the growth exponent scales linearly with $g_{a\gamma}$. We refer to this as the \textit{coherent} regime.\\
\newline
Conversely, if the amplification length is larger than the axion coherence length, the photon samples several axion domains before appreciable amplification develops. The different axion momentum components then dephase, preventing a coherent buildup of the electromagnetic field. In this limit, an effectively kinetic behavior emerges, with a growth exponent proportional to $g_{a\gamma}^2$. We refer to this as the \textit{incoherent} regime.\\
\newline
The axion coherence length has previously been identified as controlling whether exponential growth with $s\propto g_{a\gamma}$ can develop in Refs.~\cite{Hertzberg_2018, Arza2019, Arza_2020, Levkov:2020txo}. 
Refs.~\cite{Arza2019, Levkov:2020txo} identified the role of axion coherence in determining whether the scaling of the growth exponent with $g_{a\gamma}$ is linear or quadratic. Moreover, the two regimes have been studied in the context of axion self-interactions in Refs.~\cite{Sikivie:2009qn, Erken:2011dz}, where they are called condensed and particle kinetic regimes.\\
\newline
In this work, for the first time, we derive within a unified framework an expression for the growth rate that smoothly connects the coherent and incoherent regimes. Starting from the Maxwell equations, we show how dephasing among different axion momentum modes modifies the electromagnetic instability.\\
\newline
The paper is organized as follows. In Sec.~\ref{sec:coh}, we review the standard treatment of \textit{parametric resonance} by solving the modified Maxwell equations in the presence of a coherent, monochromatic axion field. We show that the electromagnetic instability leads to exponential growth with $s\propto g_{a\gamma}$ and discuss how a finite axion momentum modifies the resonance condition without suppressing the instability \textit{per se}, following Ref.~\cite{Arza_2020}.\\
However, if the axion constitutes all or part of the DM, it generally possesses a finite momentum dispersion, and can be approximated as an effectively coherent mode only on scales smaller than its coherence length. On larger scales, different momentum components progressively dephase. \\
In Sec.~\ref{sec:super}, thus, we consider a discrete superposition of axion momentum modes and show explicitly how this dephasing drives the transition from the coherent regime, characterized by $s\propto g_{a\gamma}$, to the incoherent regime, where $s\propto g_{a\gamma}^2$. \\
In Sec.~\ref{sec:cont}, we take the continuum limit and consider a Maxwell-Boltzmann velocity distribution representative of virialized DM. We derive an implicit expression for the photon growth exponent that correctly reproduces the coherent and incoherent asymptotic regimes and provides an approximate description of the crossover. We further argue that the qualitative structure of this result, and in particular the scaling with $g_{a\gamma}$ in the two asymptotic limits, is largely insensitive to the precise form of the chosen velocity distribution. Finally, we summarize our conclusions and discuss future directions in Sec.~\ref{sec:concl}. Appendices~\ref{app:local} and \ref{app:full} complement the main discussion.

\section{Parametric resonance in a coherent axion background}\label{sec:coh}
In this section, we review the standard treatment of parametric resonance. In the following, we treat $m_a$ and $g_{a\gamma}$ as independent parameters, with the QCD relation recovered as a special case.
Starting from the Lagrangian in Eq.~\eqref{eq:lagr} and imposing the Coulomb gauge ($\vec\nabla\cdot\vec A=0$), the modified Maxwell equations can be written as~\cite{Arza_2020}:
\begin{align}
\nabla^2A_0 =& ~ g_{a\gamma} \vec\nabla a\cdot\vec\nabla\times\vec A \label{eq:A0}
\\
\left(\partial_t^2-\nabla^2\right)\vec A+\partial_t\vec\nabla A_0 =&  -g_{a\gamma} \vec\nabla a\times(\partial_t\vec A+\vec\nabla A_0)+ g_{a\gamma} \partial_ta\vec\nabla\times\vec A \label{eq:A}
\\
\left(\partial_t^2-\nabla^2+m_a^2\right)a =&  -g_{a\gamma} (\partial_t\vec A+\vec\nabla A_0)\cdot\vec\nabla\times\vec A \enspace. \label{eq:a}
\end{align}
We consider \textit{coherent} axion fields in the following, namely fields whose phase remains well defined over the spatial and temporal scales relevant to the process under consideration. Initially, we focus on the simple case of a zero-momentum axion~\cite{Arza2019, Hertzberg_2018, Alvarez:2020}; then, we generalize the analysis to a monochromatic wave with non-zero momentum~\cite{Arza_2020}.\\
For simplicity, we restrict to one-dimensional propagation along the $z$-axis. While this is exact in the case of zero-momentum axions, it represents an approximation when the axion possesses a momentum. However, as we are interested in a non-relativistic axion, this is not expected to qualitatively affect the following discussion. We will comment more specifically on this issue in Sec.~\ref{subsubsec:comment}. Throughout the analysis, we neglect the backreaction of the amplified electromagnetic field on the axion background, leaving its inclusion for future work.

\subsection{Zero-momentum axion and the Mathieu instability}\label{sec:zeromom}

A zero-momentum axion field can be written as
\begin{equation}
    a(t) = a_0 \sin(m_a t + \phi_a), \qquad a_0 = \frac{\sqrt{2\rho_a}}{m_a},
    \label{eq:at}
\end{equation}
where $\phi_a$ denotes an arbitrary phase and the amplitude $a_0$  is related to the axion energy density $\rho_a$. For numerical estimates, we normalize $\rho_a$ to the benchmark local Galactic DM density, namely $\rho_\text{DM} = 0.4~\text{GeV}/\text{cm}^3$.\\
Rather than being merely a theoretical simplification, the form in Eq.~\eqref{eq:at} also provides an appropriate description of a virialized DM axion field when considering lengthscales smaller than its coherence length, or equivalently timescales shorter than its coherence time (see Sec.~\ref{sec:super} and Appendix~\ref{app:local} for further details). Owing to the stochastic nature of the axion field, however, different coherence patches are characterized by different values of the local density $\rho_a$. The quantity $\rho_\text{DM}$ should therefore be understood as the spatial, or equivalently temporal, average over many such patches. We discuss these aspects further in Appendix~\ref{app:local} and refer the interested reader to Refs.~\cite{Foster:2018, Centers:2021, 
Foster:2026mvs}.
These aspects do not play a central role in the present analysis, and we therefore neglect them in what follows.
\newline
Using the form in Eq.~\eqref{eq:at}, the spatial gradients of the axion field vanish trivially, and Eq.~\eqref{eq:A0} is consistently solved by $A_0 = 0$. Thus, Eq.~\eqref{eq:A} reduces to
\begin{equation}
    \left(\partial_t^2 - \nabla^2\right)\vec{A} = g_{a\gamma}\left( \partial_t a ~\vec \nabla \times \vec{A} \right).
    \label{eq:Azeromom}
\end{equation}
As discussed above, we specialize to one-dimensional propagation along the $z$-axis, which is exact for a zero-momentum axion background. For a photon propagating in the positive $z$ direction, the vector potential $\vec A$ can be decomposed into circular polarization states as~\cite{Hertzberg_2018}
\begin{equation}
    \vec{A} = \sum_{\lambda  = \pm} \vec\epsilon_\lambda A_\lambda(t)e^{ikz} + \text{c.c.},
\end{equation}
where the transverse polarization vectors, $\vec\epsilon_\lambda$, satisfy 
\begin{equation}
    i\hat{z} \times \vec\epsilon_\lambda = \lambda \vec\epsilon_\lambda, \qquad \lambda = \pm.
    \label{eq:lpm}
\end{equation}
Then, Eq.~\eqref{eq:Azeromom} becomes
\begin{equation}
    \ddot A_\lambda + \left[k^2 - \lambda g_{a\gamma} k m_a a_0 \cos (m_a t + \phi_a)\right]A_\lambda = 0,
    \label{eq:Alambda}
\end{equation}
where the two circular polarizations have decoupled. This equation  can be mapped onto the standard Mathieu equation~\cite{Kovacic:2018, Arscott:1968}
\begin{equation}
    \frac{d^2 y}{dx^2} + \left[\mathcal{A} - 2\mathcal{Q} \cos (2x)\right]y = 0,
    \label{eq:mathieu}
\end{equation}
through the substitutions
\begin{equation}
    x = \frac{m_at}{2} + \frac{\phi_a}{2}, \qquad \mathcal{A} = \frac{4k^2}{m_a^2}, \qquad \mathcal{Q} = 2\frac{g_{a\gamma}a_0\lambda k}{m_a}.
    \label{eq:sub}
\end{equation}
The axion decay process selects photon momenta close to $k\simeq m_a/2$, corresponding to the first instability band of the Mathieu equation, $\mathcal{A}\simeq1$. In the narrow-resonance regime, $|\mathcal{Q}|\ll1$, one may write
\begin{equation}
    y(x) = \alpha(x) e^{ix} + \beta(x)e^{-ix},
    \label{eq:ansatz0}
\end{equation}
where $\alpha(x), \beta(x)$ are slowly varying envelopes. 
For the systems considered here, taking $k\simeq m_a/2$ gives
\begin{equation}
    |\mathcal{Q}| = \frac{g_{a\gamma}\sqrt{2\rho_a}}{m_a} \sim 2.5\cdot10^{-16}\left(\frac{g_{a\gamma}}{10^{-10}~\text{GeV}^{-1}}\right)\left(\frac{\rho_a}{\rho_\text{DM}}\right)^{1/2}\left(\frac{10^{-6}~\text{eV}}{m_a}\right).
    \label{eq:qmin1}
\end{equation}
Thus, for ordinary Galactic DM densities and phenomenologically relevant axion-photon couplings, the system lies extremely deep in the narrow-resonance regime.
\newline
It is nevertheless useful to estimate the densities at which this approximation may cease to apply. Setting $|\mathcal{Q}|\lesssim1$ in Eq.~\eqref{eq:qmin1} gives the characteristic threshold density
\begin{equation}
    \rho_a \lesssim 6.4\cdot 10^{18}~\text{GeV}/\text{cm}^3 \left(\frac{10^{-10}~\text{GeV}^{-1}}{g_{a\gamma}}\right)^2\left(\frac{m_a}{10^{-12}~\text{eV}}\right)^2.
    \label{eq:est}
\end{equation}
Astrophysical configurations capable of reaching such large axion densities, as superradiant axion clouds around spinning black holes~\cite{Brito:2020}, are typically finite and spatially inhomogeneous, which can significantly modify the resonance dynamics~\cite{Hertzberg_2018,Ikeda:2018}. Moreover, at sufficiently large field amplitudes, axion self-interactions may substantially affect the evolution and structure of the cloud~\cite{Yoshino:2012,Baryakhtar:2021}, while sufficiently strong electromagnetic amplification can eventually induce significant backreaction on the axion field~\cite{Ikeda:2018,Spieksma:2023}. These systems also typically reside in nontrivial gravitational and plasma environments, which can modify both the axion-cloud dynamics and the photon dispersion relation, potentially altering the electromagnetic instability~\cite{Boskovic:2019,Sen:2018,Spieksma:2023}. \\
Therefore, although the homogeneous Mathieu equation can in principle be studied beyond the narrow-resonance regime, extending the present results to $|\mathcal{Q}|\gtrsim1$ in realistic astrophysical environments is not straightforward and generally requires a dedicated numerical treatment of the coupled axion-electromagnetic system.~\footnote{The normalization $m_a=10^{-12}~\mathrm{eV}$ in Eq.~\eqref{eq:est} is motivated by its relevance for superradiance around stellar-mass black holes~\cite{Brito:2017short, Brito:2017long, Brito:2020, Ellis:2026gwt}. Moreover, this scale is also comparable to the plasma frequency in sufficiently dilute environments, like the interstellar medium~\cite{Sen:2018}. This should not be regarded as necessarily representative of the environment close to the black hole, where instead plasma densities can be many orders of magnitude larger~\cite{Connors:2017,Moscibrodzka:2019}. While one would expect this to shut off the decay, instabilities may still develop in this regime for large values of $|\mathcal{Q}|$~\cite{Spieksma:2023}. For our purposes, however, Eq.~\eqref{eq:est} represents just a benchmark value of the density above which the analysis in Sec.~\ref{sec:coh} is not straightforwardly applicable.\label{foot:plasma}}\\
\newline
In the following, we restrict to the narrow-resonance regime $|\mathcal{Q}|\ll1$. In this limit, the coefficients $\alpha(x)$ and $\beta(x)$ introduced in Eq.~\eqref{eq:ansatz0} admit exponentially growing solutions proportional to $e^{\eta x}$, where
\begin{equation}
    \eta = \frac{1}{2}\sqrt{\mathcal{Q}^2 - \left(\mathcal{A}-1\right)^2}.
    \label{eq:mux}
\end{equation}
Using the definitions in Eq.~\eqref{eq:sub} and expanding around the first resonance band, this can be rewritten as
\begin{equation}
    \eta = \frac{2}{m_a}\sqrt{\kappa^2 - \frac{\delta_k^2}{4}},
\end{equation}
where we have defined 
\begin{equation}
    \kappa \equiv \frac{g_{a\gamma}a_0 m_a}{4} = \frac{g_{a\gamma}\sqrt{\rho_a}}{2\sqrt{2}}, \qquad \delta_k \equiv 2k - m_a.
    \label{eq:pars0}
\end{equation}
Returning to the original time variable, the growing mode of Eq.~\eqref{eq:Alambda} can therefore be written as
\begin{equation}
    A_\lambda(t) = e^{s_k t}~\left(\tilde\alpha ~ e^{i\frac{m_a}{2}t} + \tilde \beta e^{-i\frac{m_a}{2}t}\right),
    \label{eq:Alambdasol}
\end{equation}
where $\tilde \alpha$ and $\tilde \beta$ are constant complex coefficients, and
\begin{equation}
    s_k \equiv \frac{m_a}{2} ~\eta = \sqrt{\kappa^2 - \frac{\delta_k^2}{4}}.
    \label{eq:mut}
\end{equation}
The solution Eq.~\eqref{eq:Alambdasol} is consistent with the slowly varying envelope ansatz in Eq.~\eqref{eq:ansatz0} provided that the characteristic growth rate is much smaller than the axion oscillation frequency,
\begin{equation}
s_k \ll m_a .
\end{equation}
Equivalently, the envelope evolves on a timescale much longer than $m_a^{-1}$. Near resonance, where $s_k\simeq\kappa$, this requirement reduces, up to order one factors, to the condition $|\mathcal{Q}|\ll1$, in agreement with the narrow-resonance approximation adopted above.
At exact resonance, the \textit{pair detuning} parameter $\delta_k$ -- which we also refer to as \textit{momentum mismatch} in the following -- vanishes, and the growth rate reaches its maximum value; namely, 
\begin{equation}
    s_k = \kappa \propto g_{a\gamma}.
\end{equation}
This is the standard result for a homogeneous, zero-momentum axion background~\cite{Hertzberg_2018, Carenza_2020, Alvarez:2020} (see also e.g. Refs.~\cite{Kofman_1994, Bassett_2006, Amin_2014} for the same framework in the context of inflation and reheating): the photon-field amplitude grows as $e^{s_kt}$, while quantities quadratic in the field, such as the electromagnetic intensity or photon occupation number, grow as $e^{2s_k t}$.

\subsection{Monochromatic axions with non-zero momentum}\label{sec:nonzeromom}
We now investigate whether the exponential growth derived in the previous section relies on the assumption of a zero-momentum axion. Following Ref.~\cite{Arza_2020}, we consider a non-relativistic monochromatic axion mode with momentum $\vec p$. The generalization to an arbitrary superposition of axion momenta will be discussed in Sec.~\ref{sec:super}.
In this case, the axion field can be written as  
\begin{equation}
    a(\vec{x}, t) = a_0 \sin (\omega_a t - \vec{p}\cdot \vec{x} + \phi_a) \qquad \omega_a = \sqrt{m_a^2 + p^2}\sim m_a + \frac{p^2}{2m_a},
    \label{eq:atx}
\end{equation}
where  $p \equiv |\vec p|$ and, up to corrections suppressed in the non-relativistic limit, $a_0$ is the same amplitude introduced in Eq.~\eqref{eq:at}.
As in Sec.~\ref{sec:zeromom}, we restrict to one-dimensional propagation along the $z$-axis. Unlike in the zero-momentum case, this restriction is no longer exact. Nevertheless, for non-relativistic axions, the two photons produced in the decay are nearly back-to-back, so the one-dimensional treatment provides a good approximation. We comment on the accuracy of this approximation in more detail in Sec.~\ref{subsubsec:comment}. \\
\newline
For a finite axion momentum, spatial gradients of the axion field can no longer be neglected in Eqs.~\eqref{eq:A0}-\eqref{eq:a}. Within the one-dimensional approximation, $\vec\nabla a\parallel\hat z$, while the transverse nature of the electromagnetic waves implies $B_z = \left(\vec\nabla \times \vec A\right)_z = 0$. Therefore, Eq.~\eqref{eq:A0} still allows us to choose $A_0 = 0$. Thus, Eq.~\eqref{eq:A} becomes
\begin{equation}
    \left(\partial_t^2-\nabla^2\right)\vec A=   g_{a\gamma}\left[ \partial_ta\vec\nabla\times\vec A - \vec\nabla a\times\partial_t\vec A\right]\label{eq:Amom}.
\end{equation}
To solve Eq.~\eqref{eq:Amom}, it is convenient to decompose the electromagnetic field into positive and negative frequency components,
\begin{equation}
    \vec A = \sum_{\lambda = \pm} \vec A^+_\lambda  + \vec A^-_\lambda, \quad \vec A^-_\lambda = (\vec A^+_\lambda)^* ,
    \label{eq:Asplit}
\end{equation}
with 
\begin{equation}
    \vec{A}_\lambda^+ = \vec{\epsilon}_\lambda \alpha_k(t) e^{ikz}e^{-ikt} + \vec\epsilon_{-\lambda}\beta_q(t) e^{iqz} e^{-i|q|t},
    \label{eq:A+s}
\end{equation}
where $\alpha_k(t), \beta_q(t)$ are assumed to be slowly varying envelopes, in analogy with Eq.~\eqref{eq:ansatz0}.\\
\newline
The first term in Eq.~\eqref{eq:A+s} describes a photon with momentum $k>0$, which we take as the reference incoming \textit{signal} mode. Momentum conservation in the axion decay then fixes the momentum of the second photon to $q = p-k$. Following the terminology commonly used in the optics literature (e.g. Refs.~\cite{Hong:1985, Wang:1991, Joobeur:1996}), we refer to the latter mode as the \textit{idler} photon.  In the non-relativistic regime relevant here, and close to the resonant configuration $k\simeq m_a/2$, one has $p\ll k$ and therefore $q<0$. The idler thus propagates approximately along the negative $z$-direction, so that the two photons are nearly back-to-back. The opposite polarization labels appearing in Eq.~\eqref{eq:A+s} reflect the reversal of the direction of propagation (see Eq.~\eqref{eq:lpm}) and correspond to the same physical helicity for the two photon modes.\\
\newline
For convenience, we decompose the axion field into positive and negative frequency components as well
\begin{equation}
    a = a^+ + a^-, \qquad a^- = (a^+)^*\qquad a^+ = \frac{a_0}{2}e^{ipz - i\omega_at - i\phi_a},
    \label{eq:asplit}
\end{equation}
where we have absorbed an irrelevant phase shift into $\phi_a$ so as to express the axion field in terms of a cosine. 
In analogy to Eq.~\eqref{eq:pars0}, we define the pair detuning $\Delta_p$ as
\begin{equation}
    \Delta_p \equiv k + |q| -\omega_a = k + |p-k| -\omega_a.
    \label{eq:detmom}
\end{equation}
With these ingredients, we can now solve Eq.~\eqref{eq:Amom}. In particular, we focus on the positive frequency contribution $\vec{A^+}$. 
Then, only terms on the right-hand side of Eq.~\eqref{eq:Amom} whose spacetime dependence matches that of the modes in Eq.~\eqref{eq:A+s} can contribute resonantly.\\
\newline
To illustrate the point, consider the first term in the square brackets of Eq.~\eqref{eq:Amom}:
\begin{equation}
    \partial_t a~ \vec \nabla \times \vec A_\lambda = \left(\partial_t a^+ + \partial_t a^-\right)\left(\nabla \times \vec A^+_\lambda + \nabla \times \vec A_\lambda^-\right).
\end{equation}
Keeping only the corresponding phase dependence, the four contributions are
\begin{equation}
\begin{split}
    &\partial_t a^+ ~\vec \nabla \times \vec A_\lambda^+ \sim \left(e^{i(p+k)z - i\phi_a - i(\omega_a + k)t} + e^{i(p+q)z - i\phi_a - i(\omega_a + |q|)t}\right)\\
    &\partial_t a^+~ \vec \nabla \times \vec A_\lambda^- \sim \left(e^{iqz - i\phi_a - i|q|t +i\Delta_p t} + e^{ikz - i\phi_a - ikt + i\Delta_p t}\right)\\
    &\partial_t a^- ~\vec \nabla \times \vec A_\lambda^+ \sim \left(e^{-iqz +i\phi_a + i|q|t -i\Delta_p t} + e^{-ikz + i\phi_a + ikt - i\Delta_p t}\right)\\
    &\partial_t a^-~\vec \nabla \times \vec A_\lambda^- \sim \left(e^{-i(p+k)z  +i\phi_a + i(\omega_a + k)t} + e^{-i(p+q)z + i\phi_a + i(\omega_a + |q|)t}\right).
    \label{eq:piece}
\end{split}
\end{equation}
Focusing for simplicity on exact resonance $\Delta_p = 0$, only the second line has the same phase structure as the positive-frequency modes in Eq.~\eqref{eq:A+s}. The third line is analogously resonant with $\vec A_\lambda^-$, while the remaining terms oscillate rapidly and average out. Restricting to the resonant contribution in the positive-frequency sector therefore gives:
\begin{equation}
    \left(\partial_t^2-\nabla^2\right)\vec A_\lambda^+=   g_{a\gamma}\left[ \partial_t a^+\vec\nabla\times\vec A_\lambda^- - \vec\nabla a^+\times\partial_t\vec A^-_\lambda\right]\label{eq:A+mom}.
\end{equation}
Within the slowly varying envelope approximation, Eq.~\eqref{eq:A+mom} yields the coupled equations
\begin{align}
        &\dot\alpha_k(t) = \lambda\frac{g_{a\gamma}a_0}{4}\frac{|q|}{k}(\omega_a + p) \beta_q^*(t) e^{-i\phi_a}e^{i\Delta_p t} \equiv \kappa_\alpha  \beta_q^*(t) e^{-i\phi_a}e^{i\Delta_p t}  \label{eq:alpha}\\
        &\dot\beta^*_q(t) = \lambda\frac{g_{a\gamma}a_0}{4}\frac{k}{|q|}(\omega_a - p) \alpha_k(t) e^{i\phi_a}e^{-i\Delta_p t}\equiv \kappa_\beta\alpha_k(t) e^{i\phi_a}e^{-i\Delta_p t},
        \label{eq:betas}
\end{align}
where we have introduced $\kappa_\alpha, \kappa_\beta$ for convenience. In the following, we leave the time dependence of $\alpha_k$ and $\beta_q$ implicit for notational simplicity.\\
\newline
It is convenient to remove the asymmetry between the two equations through the redefinitions
\begin{equation}
    \alpha_k = \alpha_k^\text{sym}\sqrt{|\kappa_\alpha|} e^{i\Delta_p t/2} \qquad \beta_q = \lambda \beta_q^\text{sym}\sqrt{|\kappa_\beta|} e^{i\Delta_p t/2}.
\end{equation}
The system of Eqs.~\eqref{eq:alpha}-\eqref{eq:betas} then becomes: 
\begin{align}
        &\dot\alpha_k^\text{sym} = -i\frac{\Delta_p}{2} \alpha_k^\text{sym} +  \kappa  e^{-i\phi_a} \left(\beta^\text{sym}_q\right)^*   \label{eq:alphasym}\\
        &\left(\dot \beta^\text{sym}_q\right)^* = \kappa  e^{i\phi_a} \alpha_k^\text{sym} + i\frac{\Delta_p}{2} \left( \beta^\text{sym}_q\right)^*,
        \label{eq:betassym}
\end{align}
where now both of the equations depend on the common coupling
\begin{equation}
    \kappa \equiv \sqrt{|\kappa_\alpha||\kappa_\beta|} = \frac{g_{a\gamma}a_0 m_a}{4} = \frac{g_{a\gamma}\sqrt{\rho_a}}{2\sqrt{2}}.
    \label{eq:k}
\end{equation}
Remarkably (and expectedly), this is exactly equal to the zero-momentum growth parameter $\kappa$ defined in Eq.~\eqref{eq:pars0}, motivating the use of the same notation. The cancellation of the explicit momentum dependence follows directly from the axion dispersion relation, $\omega_a^2-p^2=m_a^2$.\\
\newline
Indeed, adopting the ansatz $\left(\alpha_k^\text{sym}, \beta_q^\text{sym} \right) \sim e^{s_k^pt}$, in analogy with the zero-momentum analysis preceding Eq.~\eqref{eq:mux}, the growing eigenvalue of the system in Eqs.~\eqref{eq:alphasym}-\eqref{eq:betassym} is characterized by
\begin{equation}
    s_k^p = \sqrt{\kappa^2- \frac{\Delta_p^2}{4}} .
    \label{eq:mutp}
\end{equation}
This expression is formally identical to the zero-momentum result in Eq.~\eqref{eq:mut}, with the only difference that the detuning $\delta_k$ is replaced by its finite-momentum counterpart $\Delta_p$. In particular, Eq.~\eqref{eq:mutp} reduces exactly to Eq.~\eqref{eq:mut} once $p=0$, as expected. \\
\newline
For completeness, we can write the growing solution of Eq.~\eqref{eq:A+mom} in analogy to Eq.~\eqref{eq:Alambdasol} as 
\begin{equation}
    \vec{A}^+_\lambda =   e^{s_k^p t} e^{i\frac{\Delta_p}{2}t}\left(\vec\epsilon_\lambda\tilde \alpha_k e^{ikz} e^{-ikt}+ \vec \epsilon_{-\lambda}\tilde\beta_q e^{iqz} e^{-i|q|t}\right),
    \label{eq:A+lambdasolmom}
\end{equation}
where $\tilde \alpha_k, \tilde \beta_q$ are  constant complex coefficients fixed by initial conditions. Adding the negative-frequency component and taking the limit $p\to0$, Eq.~\eqref{eq:A+lambdasolmom} reproduces the zero-momentum solution in Eq.~\eqref{eq:Alambdasol}, together with its complex conjugate.
As in the zero-momentum case, the validity of this solution relies on the slowly varying envelope approximation. The characteristic envelope evolution rate must remain much smaller than the fast oscillation frequencies, $k,|q|\sim m_a/2$, which again reduces, up to factors of order unity, to the narrow-resonance condition $\mathcal{Q}\ll1$ introduced in Sec.~\ref{sec:zeromom}.\\
\newline
We therefore conclude that allowing the axion field to have non-zero, non-relativistic momentum does not  qualitatively change the exponential amplification of the photon field. The maximum growth rate remains identical to that of the zero-momentum case,
\begin{equation}
s_k^p(\Delta_p=0)=\kappa\propto g_{a\gamma},
\end{equation}
while the effect of the axion momentum is entirely encoded in the modified resonance condition through the pair detuning $\Delta_p$.

\subsubsection{Accuracy of the one-dimensional approximation}\label{subsubsec:comment}

Before proceeding, let us comment on the validity of the one-dimensional approximation adopted so far. 
In three dimensions, the momentum of the \textit{signal} photon can be written as $\vec k = k \hat n$, where $\hat n$ denotes an arbitrary propagation direction. We can then decompose the axion momentum as $\vec p = p_\parallel \hat n + \vec p_\perp$, where $\vec p_\perp$ is transverse to $\hat n$. Thus, the magnitude of the \textit{idler} photon momentum can be expanded as
\begin{equation}
    |\vec q\,| = |\vec p -\vec{k}| = k-p_\parallel + \frac{p_\perp^2}{2k}+...
\end{equation}
Using this result, together with the expansion of $\omega_a$ in Eq.~\eqref{eq:atx}, the pair detuning parameter defined in Eq.~\eqref{eq:detmom} becomes:
\begin{equation}
    \Delta_p = 2k- m_a -p_\parallel + \frac{p_\perp^2}{2k} - \frac{p^2}{2m_a} + ...
    \label{eq:deltamomexp}
\end{equation}
Equation~\eqref{eq:deltamomexp} shows that the axion momentum enters the detuning at linear order only through its longitudinal component, while the transverse momentum first appears at quadratic order. This justifies the one-dimensional treatment as the leading-order approximation in the non-relativistic expansion. Physically, this is consistent with the fact that the decay of a non-relativistic axion produces two photons that are nearly back-to-back, with deviations from perfect collinearity suppressed by the axion velocity. In the following, we will label $p_\parallel$ as $p$, to avoid cluttering the notation. However, it is understood that we are referring to the longitudinal component of the axion momentum.

\section{From coherent to incoherent amplification: finite axion coherence}\label{sec:super}
In Sec.~\ref{sec:coh}, we showed that a generic monochromatic axion wave, including the zero-momentum limit, can trigger a parametric resonance leading to an exponential amplification of the photon field. The corresponding growth exponent scales linearly with the axion-photon coupling, namely $s\propto g_{a\gamma}$ (see Eqs.~\eqref{eq:mut} and \eqref{eq:mutp}), in agreement with previous studies of coherent axion backgrounds~\cite{Hertzberg_2018, Carenza_2020, Arza_2020, Alvarez:2020}.\\
\newline
At first sight, this linear dependence may appear surprising from a particle-based perspective. Indeed, in a kinetic description based on the Boltzmann equation, the evolution of phase-space occupation numbers is governed by collision terms depending on the squared matrix element of the process under consideration~\cite{Kolb:1990vq}. Since the amplitude for $a\to\gamma\gamma$ is proportional to $g_{a\gamma}$, the corresponding rate scales as $g_{a\gamma}^2$. Therefore, if an exponential growth develops within such a description, we would expect it to be controlled by $g_{a\gamma}^2$, not $g_{a\gamma}$. Indeed, in a companion work~\cite{Smarra:2026}, we study stimulated axion decay using a Boltzmann description in the context of DM spikes~\cite{Gondolo_1999, Ullio_2001, Bertone_2005, Shen:2023kkm, Lacroix:2018zmg}, finding an exponentially growing photon population with a characteristic rate proportional to $g_{a\gamma}^2$.\\
\newline
The crucial difference between the two descriptions lies in the information they retain about the phases of the underlying fields. The classical Maxwell equations evolve the electromagnetic and axion field amplitudes themselves, and therefore keep track of their relative phases. A standard Boltzmann equation, instead, evolves only phase-space occupation numbers and does not retain the off-diagonal phase correlations between different momentum modes~\cite{Sigl:1993,Raffelt:1993,Berges:2006,Vlasenko:2014}.
This suggests that the linear behavior found above is intrinsically tied to the coherence of the axion field.\\
\newline
The monochromatic configurations considered in Eqs.~\eqref{eq:at} and \eqref{eq:atx} are perfectly coherent. A realistic axion DM field, however, is a superposition of momentum modes with a finite velocity dispersion $\sigma_v$. As a consequence, its phase remains correlated only over a finite coherence length,
\begin{equation}
L_c\sim\frac{1}{m_a\sigma_v}\sim200~\text{m}\left(\frac{10^{-6}~\text{eV}}{m_a}\right)\left(\frac{10^{-3}}{\sigma_v}\right),
\label{eq:Lc}
\end{equation}
and over a corresponding coherence time,
\begin{equation}
\tau_c\sim\frac{1}{m_a\sigma_v^2}\sim6.6\times10^{-4}~\text{s}
\left(\frac{10^{-6}~\text{eV}}{m_a}\right)\left(\frac{10^{-3}}{\sigma_v}\right)^2 .
\label{eq:tc}
\end{equation}
Here $\sigma_v\sim10^{-3}$ is a benchmark value representative of the velocity dispersion of the virialized Galactic DM halo~\cite{ParticleDataGroup:2024cfk}. On spatial scales smaller than $L_c$, or temporal scales shorter than $\tau_c$, the axion field can be treated locally as a coherent oscillating background with approximately fixed amplitude and phase, as in Eq.~\eqref{eq:at}, with these quantities varying stochastically from one coherence patch to another. We discuss this picture further in Appendix~\ref{app:local}; see also Refs.~\cite{Foster:2018,Centers:2021,Foster:2026mvs}.\\
The coherence time in Eq.~\eqref{eq:tc} can be understood parametrically as the time required for an axion moving with velocity $\sim\sigma_v$ to traverse one coherence length. From the perspective of a photon propagating at the speed of light, the relevant timescale is instead
\begin{equation}
\tau_c^{\rm cross}\sim\frac{L_c}{c}\sim\frac{1}{m_a\sigma_v},
\end{equation}
where in the last expression we have set $c=1$. Since $\tau_c^{\rm cross}\ll\tau_c$ for non-relativistic axions, the coherence properties of a given patch change negligibly while the photon traverses it.\\
\newline
This observation suggests a simple physical distinction. Let $s$ denote the coherent growth rate obtained from the analysis of Sec.~\ref{sec:coh}, suppressing the momentum labels for notational simplicity. If $s^{-1}\lesssim L_c$, the photon experiences significant amplification before leaving a single coherence patch. The coherent-field analysis of Sec.~\ref{sec:coh} is then applicable, and the growth exponent scales linearly with $g_{a\gamma}$.\\
Conversely, if $s^{-1}\gtrsim L_c$, the photon traverses many effectively uncorrelated coherence domains before substantial amplification can develop. The coherent buildup is then disrupted by the relative dephasing of the different axion momentum modes, and the evolution is expected to cross over toward an incoherent, kinetic description in which the characteristic interaction rate scales as $g_{a\gamma}^2$~\cite{Kolb:1990vq, Smarra:2026}.\\
\newline
It is therefore useful to introduce the dimensionless parameter
\begin{equation}
\Theta\equiv\frac{s}{m_a\sigma_v},
\label{eq:Theta}
\end{equation}
which compares the coherent growth rate with the inverse photon crossing time of an axion coherence domain. Parametrically, $\Theta\gtrsim1$ corresponds to the coherent regime, in which appreciable exponential amplification can develop within a single coherence patch and the growth exponent scales as $g_{a\gamma}$. For $\Theta\lesssim1$, instead, the photon samples multiple incoherent patches before substantial growth occurs, and the evolution is expected to cross over toward the incoherent, $g_{a\gamma}^2$-controlled regime.
The remainder of this section is devoted to testing this intuitive picture and quantifying the transition between the coherent and incoherent regimes.

\subsection{Discrete superposition of axion momentum modes}\label{sec:discrmom}
We now consider an axion field composed of a discrete superposition of momentum modes:
\begin{equation}
    a(z,t) = \sum_j a_j \cos (\omega_j t - p_jz + \phi_j), \qquad \omega_j = \sqrt{m_a^2 + p_j^2} \sim m_a + \frac{p_j^2}{2m_a},
    \label{eq:supmom}
\end{equation}
where, as throughout the previous discussion, we restrict to one-dimensional propagation; see Sec.~\ref{subsubsec:comment} for a discussion of this approximation. This setup provides a simple framework in which the loss of coherence can be followed explicitly, allowing us to study how the interference among different axion modes modifies the exponential photon amplification derived in Sec.~\ref{sec:coh}.
For convenience, we again decompose each axion mode into positive and  negative frequency components:
\begin{equation}
    a = \sum_j a^+_j + a_j^-, \qquad a^-_j = (a^+_j)^* ,\qquad a_j^+ = \frac{a_j}{2}e^{-i\omega_j t + ip_j z - i\phi_j}.
    \label{eq:axcomp}
\end{equation}
For each axion momentum component, we define the corresponding idler momentum and pair detuning as
\begin{equation}
    q_j = p_j -k, \qquad \Delta_j = k + |q_j| - \omega_j.
\end{equation}
Conceptually, the problem is a straightforward generalization of the monochromatic case discussed in Sec.~\ref{sec:nonzeromom}: we just need to keep track of different momentum modes. We therefore continue to solve Eq.~\eqref{eq:Amom} and decompose the vector potential $\vec A$  as in Eq.~\eqref{eq:Asplit}. However, its positive-frequency component is now taken to have the form
\begin{equation}
    \vec{A}_\lambda^+ = \vec{\epsilon}_\lambda \alpha_k(t) e^{ikz}e^{-ikt} + \sum_j\vec\epsilon_{-\lambda}\beta_j(t) e^{iq_jz} e^{-i|q_j|t},
    \label{eq:A+}
\end{equation}
where $\alpha_k$ denotes the reference signal mode, while each axion momentum $p_j$ is associated with an idler mode of momentum $q_j$. \\
\newline
A caveat is in order. Strictly speaking, the ansatz in Eq.~\eqref{eq:A+} is not closed under the full Maxwell evolution: additional photon modes with momenta close to, but different from, the reference momentum $k$ are in general generated dynamically. As we show in Appendix~\ref{app:full}, this simplified treatment nevertheless reproduces the correct coherent and incoherent asymptotic limits. In the intermediate regime, the additional mode couplings can lead to quantitative corrections to the crossover, although we do not expect them to modify its qualitative behavior. We therefore retain the simplified ansatz in Eq.~\eqref{eq:A+} in the following.\\
\newline
Proceeding as in Sec.~\ref{sec:nonzeromom} and retaining only the resonant contributions in Eq.~\eqref{eq:Amom}, we obtain
\begin{align}
        &\dot\alpha_k = \sum_j\lambda\frac{g_{a\gamma}a_j}{4}\frac{|q_j|}{k}(\omega_j + p_j) \beta_j^* e^{-i\phi_j}e^{i\Delta_j t} \equiv \sum_j\kappa_{\alpha j}  \beta_j^* e^{-i\phi_j}e^{i\Delta_j t}  \label{eq:alphatot}\\
        &\dot\beta^*_j = \lambda\frac{g_{a\gamma}a_j}{4}\frac{k}{|q_j|}(\omega_j - p_j) \alpha_k e^{i\phi_j}e^{-i\Delta_j t}\equiv \kappa_{\beta j}\alpha_k e^{i\phi_j}e^{-i\Delta_j t}.
        \label{eq:betasj}
\end{align}
As a consistency check, when the axion field contains a single momentum component, this system immediately reduces to Eqs.~\eqref{eq:alpha}-\eqref{eq:betas}. The equations can be symmetrized through the redefinitions
\begin{equation}
    \alpha_k = \alpha_k^\text{sym },\qquad \beta_j = \lambda\beta_j^\text{sym} \sqrt{\frac{|\kappa_{\beta j}|}{|\kappa_{\alpha j}|}},
\end{equation}
so that
\begin{align}
        &\dot\alpha_k^\text{sym} =  \sum_j\kappa_{j}  \left(\beta_j^\text{sym}\right)^* e^{-i\phi_j}e^{i\Delta_j t}  \label{eq:alphatotsym}\\
        &\left(\dot\beta_j^\text{sym}\right)^*  = \kappa_{j}\alpha_k^\text{sym} e^{i\phi_j}e^{-i\Delta_j t},
        \label{eq:betasjsym}
\end{align}
where, in analogy with Eq.~\eqref{eq:k}, we have defined
\begin{equation}
    \kappa_j \equiv \sqrt{|\kappa_{\alpha j}||\kappa_{\beta j}|} = \frac{g_{a\gamma}a_j m_a}{4} = \frac{g_{a\gamma}\sqrt{\rho_j}}{2\sqrt{2}}, \qquad \kappa = \sqrt{\sum_j \kappa_j^2},
    \label{eq:kj}
\end{equation}
with $\rho_j$ the averaged energy density in the axion $p_j$ momentum mode (see Appendix~\ref{app:local} for more details).\\
\newline
Integrating explicitly Eq.~\eqref{eq:betasjsym} and plugging it back into Eq.~\eqref{eq:alphatotsym}, we obtain:
\begin{equation}
    \begin{split}
        \dot \alpha_k^\text{sym}(t) &= \sum_j \kappa_j e^{-i\phi_j} e^{i\Delta_j t} \left(\beta_j^\text{sym}(0)\right)^* + \sum_j \kappa_j^2 \int_0^t dt' \alpha_k^\text{sym}(t') e^{i\Delta_j (t-t')} \\
        &\equiv S(t) + \int_0^t d\tau ~K(\tau)\alpha_k^\text{sym}(t-\tau),
    \end{split}
    \label{eq:volt}
\end{equation}
where in the last line we have introduced $\tau\equiv t-t'$. Equation~\eqref{eq:volt} is a linear Volterra integro-differential equation. The first term on the right hand side, $S(t)$, acts as a source determined by the initial occupation of the idler modes, and can be conveniently set to zero. Then, the dynamics is entirely controlled by the second term, which is a convolution depending on the \textit{kernel}
\begin{equation}
    K(\tau) \equiv \sum_j \kappa_j^2 e^{i\Delta_j\tau}.
    \label{eq:kernel}
\end{equation}
The kernel encodes the cumulative contribution of the different axion momentum components and, crucially, their relative dephasing through the factors $e^{i\Delta_j\tau}$. As we show below, it is this dephasing that governs the transition away from the fully coherent exponential-growth regime.\\
\newline
Before proceeding to a detailed analysis of Eq.~\eqref{eq:volt}, which we postpone to Sec.~\ref{sec:cont}, it is useful to develop some intuition for the two limiting regimes.\\
Consider first a single axion mode whose associated signal-idler pair is exactly resonant, so that $\kappa_1=\kappa$ and $\Delta_1=0$. Equation~\eqref{eq:volt} then reduces to 
\begin{equation}
    \dot\alpha_k^\text{sym } (t) = \kappa^2\int_0^t dt' \alpha_k^\text{sym}(t') \longrightarrow \ddot\alpha_k^\text{sym}(t) = \kappa^2 \alpha_k^\text{sym}(t),
\end{equation}
which is precisely the $\Delta_p  =0$ limit of Eqs.~\eqref{eq:alphasym}-\eqref{eq:betassym}. Thus, $\alpha_k^\text{sym}(t)$ admits an exponentially growing solution, with exponent $\kappa \propto g_{a\gamma}$, recovering the coherent result discussed above.\\
\newline
The situation changes when the axion field contains many momentum modes with comparable amplitudes but different detunings. In this case, the different phase factors entering the kernel in Eq.~\eqref{eq:kernel} progressively dephase. Denoting by $\sigma_\Delta$ the characteristic width of the detuning distribution, the kernel remains coherent only for time separations of order $\tau_\text{coh} \sim \sigma_\Delta^{-1}$.  Beyond this timescale, the relative phases accumulated by the different momentum components become of order unity, and their contributions to the kernel largely cancel.\\
It is useful to relate this timescale to the coherence scales introduced at the beginning of Sec.~\ref{sec:super}. For non-relativistic axions, the detuning can be expanded at leading order as
\begin{equation}
\Delta_j \simeq \delta_k-p_j+\cdots,
\label{eq:Deltaexp}
\end{equation}
with the momentum mismatch $\delta_k$ defined in Eq.~\eqref{eq:pars0}.
The spread of the detuning distribution is therefore controlled by the axion momentum dispersion. For a velocity distribution characterized by the dispersion $\sigma_v$, one finds $\sigma_\Delta \sim m_a \sigma_v$. The timescale $\tau_\text{coh}$ then coincides parametrically with the photon crossing time $\tau^\text{cross}_c$, and should not be confused with the intrinsic temporal coherence of the axion field $\tau_c$, which is parametrically longer by a factor $\sim c/v$.\\
\newline
To illustrate this behavior schematically, let us focus on the resonant case, $\delta_k=0$, and assume that the axion density is distributed over a sufficiently large number of momentum modes with comparable weights, such that no single contribution dominates the sum in Eq.~\eqref{eq:kernel}. Since we are interested here in the exponential amplification of $\alpha_k^{\rm sym}$, we focus on the real part of the kernel, which may be approximated as
\begin{equation}
    K(\tau) \sim \begin{cases}
        &\kappa^2\qquad \tau \lesssim \tau_\text{coh}\\
        &0\qquad \,\,\,\tau>\tau_\text{coh}
    \end{cases}\longrightarrow \int_0^\infty d\tau~K(\tau) \sim \kappa^2 \tau_\text{coh},
    \label{eq:schem}
\end{equation}
where the sharp cutoff is only meant to capture qualitatively the suppression induced by dephasing.
If $\alpha_k^{\rm sym}(t)$ varies slowly over the kernel correlation time $\tau_\text{coh}$, it can be taken outside the convolution in Eq.~\eqref{eq:volt}, yielding
\begin{equation}
    \dot\alpha_k^\text{sym} (t) \sim \kappa^2\tau_\text{coh} \alpha_k^\text{sym} (t).
    \label{eq:slow}
\end{equation}
The resulting evolution is again exponential
\begin{equation}
    \alpha_k^\text{sym}(t) \sim e^{\kappa^2\tau_\text{coh} ~t} ,
    \label{eq:qual}
\end{equation}
but the growth exponent now scales as  $\kappa^2\tau_\text{coh} \propto g_{a\gamma}^2$.\\
\newline
This reproduces the intuitive picture developed at the beginning of Sec.~\ref{sec:super}. If the characteristic amplification timescale of $\alpha_k^{\rm sym}$ is shorter than the photon crossing time of an axion coherence domain, significant growth can occur before the different axion momentum modes dephase, and the coherent $g_{a\gamma}$ scaling is recovered. Conversely, if the amplification timescale is longer than the dephasing time $\tau_{\rm coh}\sim\tau_c^{\rm cross}$, the photon probes several effectively incoherent contributions before substantial amplification develops, and the effective growth rate crosses over to the $g_{a\gamma}^2$ scaling.\\
\newline
Before proceeding, it is useful to compare the result in Eq.~\eqref{eq:qual} with the treatment developed in Sec.~III of Ref.~\cite{Arza_2020}. That work also considers a superposition of axion momentum modes and investigates whether it can sustain a coherent, Mathieu-like instability. When $\kappa \gtrsim \sigma_\Delta$, all the axion momentum components add up coherently, and Ref.~\cite{Arza_2020} recovers the standard Mathieu-like growth, in agreement with our discussion. When $\kappa \lesssim \sigma_\Delta$, instead, only a restricted portion of the axion momentum distribution can participate coherently in the Mathieu-like instability.
In this regime, Ref.~\cite{Arza_2020} enters the so-called \textit{singular} regime, in which the asymptotic coherent growth exponent depends on the momentum resolution, effectively measuring the fraction of the axion field that can participate in the coherent buildup. Consequently, this singular exponent vanishes in the continuum limit.\\
At first sight, this may seem in tension with Eq.~\eqref{eq:qual}, which instead predicts a discretization-independent and non-vanishing asymptotic growth exponent. This apparent discrepancy can be resolved by noticing that the two approaches characterize different amplification mechanisms. Ref.~\cite{Arza_2020} analyzes whether a coherent instability can still be sustained when the axion momentum distribution becomes broader than the characteristic Mathieu-like resonance width. The fact that their singular exponent depends on the momentum discretization reflects the decreasing fraction of the axion field that can remain phase-coherent within a given unstable mode. Consistently, this contribution remains linear in $g_{a\gamma}$, as expected for a Mathieu-like picture.\\
Here, by contrast, we ask what remains once this coherent buildup is lost. In the incoherent regime, different axion momentum components contribute with different detunings and phases and do no longer add coherently at the photon amplitude level. Nevertheless, their cumulative effect survives, and is encoded in the kernel. In this sense, Ref.~\cite{Arza_2020} characterizes the disappearance of the coherent $\mathcal O(g_{a\gamma})$ contribution, while the present treatment shows that a weaker incoherent $\mathcal O(g_{a\gamma}^2)$ gain can remain. This interpretation is also consistent with our Boltzmann-inspired analysis in Ref.~\cite{Smarra:2026}, where the corresponding growth exponent scales as $s\propto g_{a\gamma}^2$.

\section{The continuum limit and the coherent-to-incoherent crossover}\label{sec:cont}

In the previous section, we derived the evolution equation for a discrete superposition of axion momentum modes and showed that the effect of the momentum distribution is encoded in the kernel $K(\tau)$ in Eq.~\eqref{eq:kernel}. We also argued that the characteristic timescale over which this kernel loses coherence is set by the photon crossing time of an axion coherence domain. This naturally distinguishes two regimes: if the amplification timescale is shorter than the photon crossing time, coherent growth can develop within a single domain; if instead amplification is slower, dephasing becomes important before substantial growth takes place. We now extend this framework to a continuous axion velocity distribution and determine quantitatively how the growth rate interpolates between these two regimes.\\
\newline
By using the property $\int d\Delta'~\delta(\Delta'-\Delta_j) = 1$, we may rewrite the kernel in Eq.~\eqref{eq:kernel} as
\begin{equation}
    K(\tau) = \kappa^2\int d\Delta' \left[\sum_j \frac{\kappa_j^2}{\kappa^2}\delta (\Delta' - \Delta_j)\right]e^{i\Delta' \tau}\equiv \kappa^2\int d\Delta' P(\Delta') e^{i\Delta '\tau} ,
    \label{eq:Pdelta}
\end{equation}
where we have defined the \textit{weighted detuning distribution} $P(\Delta')$. Before further specifying its form, it is useful to recover the incoherent result discussed heuristically in the previous section from a more general standpoint. If $\alpha_k^\text{sym}$ varies negligibly over the correlation time of the kernel, we may approximate $\alpha_k^\text{sym}(t-\tau)\sim \alpha_k^\text{sym}(t)$ inside Eq.~\eqref{eq:volt} and extend the upper integration limit to infinity. Using the distributional identity
\begin{equation}
    \int_0^\infty d\tau ~e^{i\Delta \tau} = \pi\delta (\Delta) + i ~\text{PV} \frac{1}{\Delta},
\end{equation}
where $\text{PV}$ is the principal value, we obtain
\begin{equation}
    \dot\alpha_k^\text{sym} (t) =\alpha_k^\text{sym}(t) \int_0^\infty d\tau~K(\tau)  = \alpha_k^\text{sym}(t)\kappa^2\left[ \pi P(0) + i\text{PV} \int d\Delta' \frac{P(\Delta')}{\Delta'}\right].
    \label{eq:mark}
\end{equation}
For an even detuning distribution, the principal-value contribution vanishes, leaving
\begin{equation}
    \dot\alpha_k^\text{sym} (t)  = \pi P(0)\kappa^2~\alpha_k^\text{sym}(t) \longrightarrow \alpha_k^\text{sym} (t)\sim e^{\pi P(0)\kappa^2 ~t}.
    \label{eq:mark2}
\end{equation}
Notice that extending the time integral to infinity does not require $\alpha_k^\text{sym}$ to vary slowly over arbitrarily long times. Rather, the approximation only requires it to vary slowly over the finite timescale in which $K(\tau)$ has appreciable support. As discussed in Sec.~\ref{sec:discrmom}, this timescale is parametrically equal to the photon crossing time of one axion coherence domain.\\
\newline
We now take the continuum limit explicitly. From Eq.~\eqref{eq:kj}, we have
\begin{equation}
    \kappa_j ^2 \propto \rho_j \ \propto \int_{v_j}^{v_j + \delta v_j} dv~ f(v) \sim f(v_j)\delta v_j,
    \label{eq:dv}
\end{equation}
where $f(v)$ is the normalized one-dimensional velocity distribution. Since the overall normalization cancels in the ratio $\kappa_j^2/\kappa^2$, the continuum limit of $P(\Delta')$ becomes
\begin{equation}
    P(\Delta') = \int dv ~f(v) \delta \left(\Delta' - \Delta(v)\right).
    \label{eq:contp}
\end{equation}
In analogy to Eq.~\eqref{eq:Deltaexp}, we can write
\begin{equation}
    \Delta(v) \sim  \delta_k - m_av.
    \label{eq:Deltav}
\end{equation}
As we will see, sufficiently large values of $|\delta_k|$ will correspond to strongly suppressed amplification.

\begin{figure}[t]
  \centering
  \includegraphics[width=0.98\columnwidth]{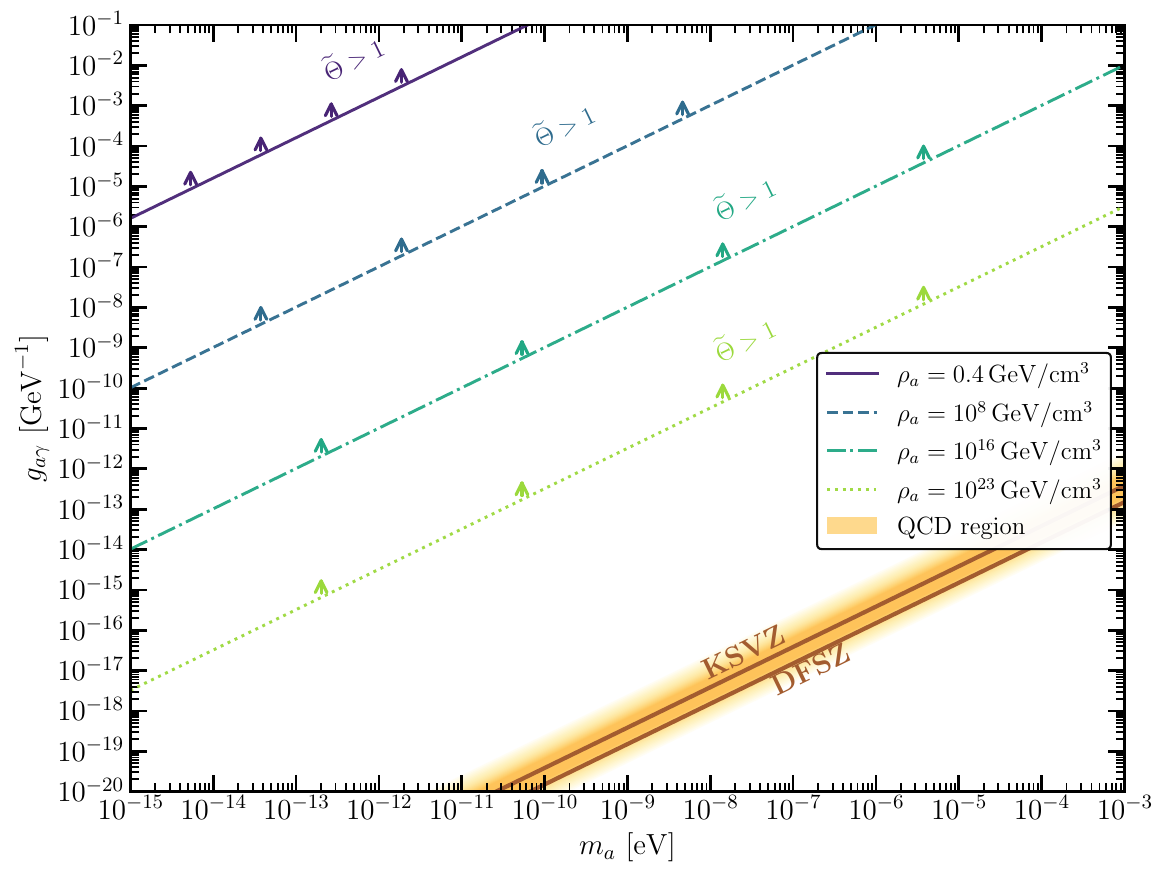}
  \caption{Division between the coherent and incoherent regimes, namely $\tilde{\Theta}>1$ and $\tilde{\Theta}\leq 1$, for different values of the axion density $\rho_a$, keeping the velocity dispersion $\sigma_v$ fixed to the benchmark value in Eq.~\eqref{eq:tt}.  We also show the QCD axion band and the KSVZ~\cite{Kim:1979, SHIFMAN1980493} and DFSZ~\cite{DINE1981199, Zhitnitsky:1980tq} benchmark
    lines, produced using the ``AxionLimits" GitHub repository~\cite{AxionLimits}.  }
  \label{fig:theta_regions}
\end{figure}

\begin{figure}[t]
  \centering
  \includegraphics[width=0.98\columnwidth]{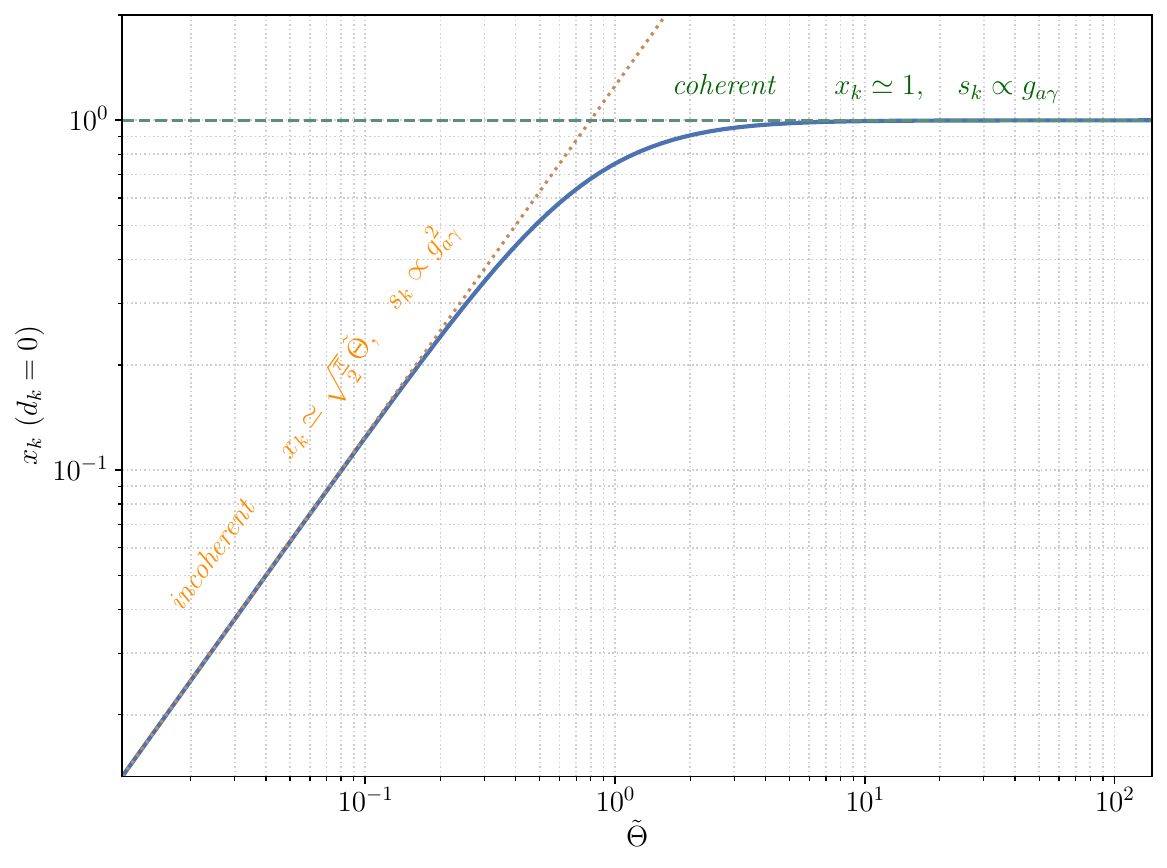}
  \caption{Dimensionless growth exponent $x_k$ as a function of $\tilde\Theta$, evaluated at exact resonance, $d_k=0$. The solid curve shows the numerical solution of Eq.~\eqref{eq:implicitadim}, while the asymptotic expressions in Eqs.~\eqref{eq:inclim} and \eqref{eq:xc} describe the incoherent ($\tilde\Theta\ll1$) and coherent ($\tilde\Theta\gg1$) limits, respectively. The curve smoothly interpolates between the $s_k\propto g_{a\gamma}^2$ and $s_k\propto g_{a\gamma}$ regimes.}
  \label{fig:growthrate}
\end{figure}

\begin{figure}[t]
  \centering
  \includegraphics[width=0.98\columnwidth]{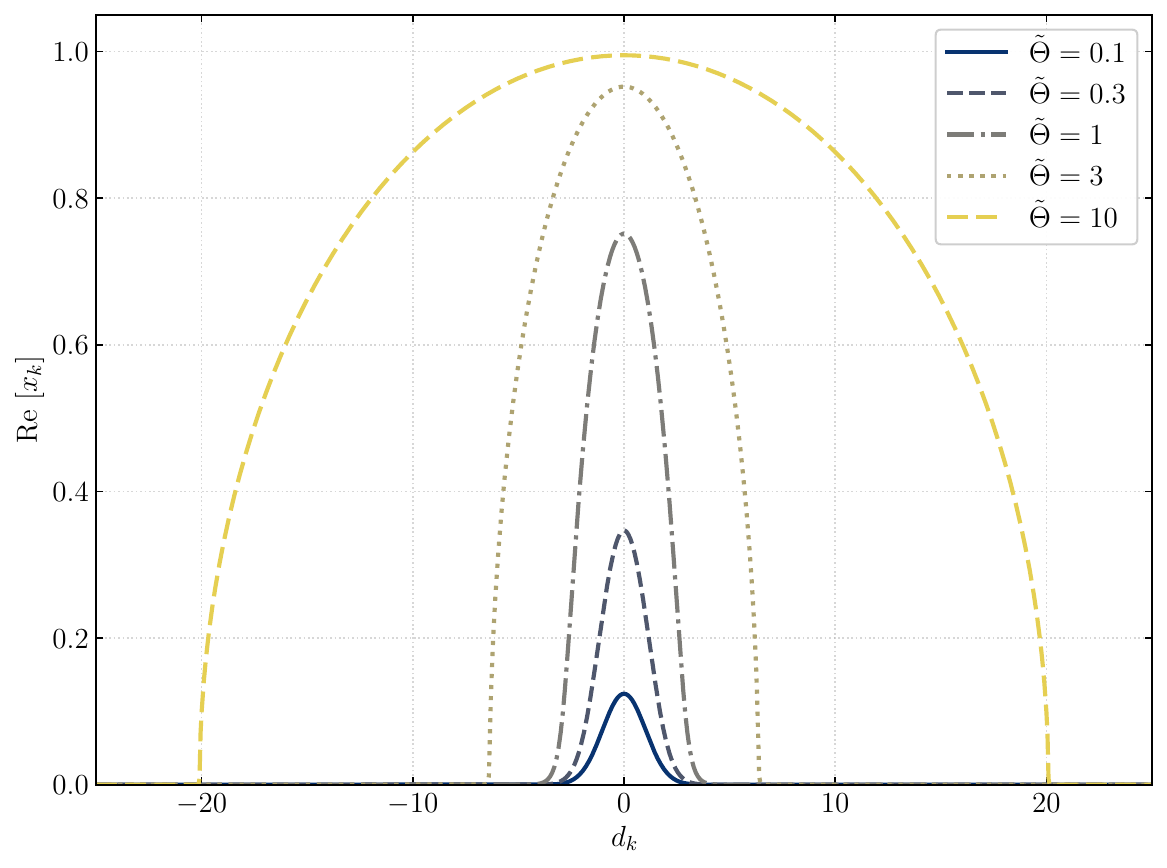}
  \caption{Real part of the dimensionless growth exponent, $\text{Re}\,[x_k]$, as a function of the normalized momentum mismatch $d_k$, obtained by solving Eq.~\eqref{eq:implicitadim} for representative values of $\tilde\Theta$. Increasing $\tilde\Theta$ continuously deforms the resonance profile from the Gaussian behavior characteristic of the incoherent regime to the square-root profile of the coherent limit. Both the peak growth rate and the resonance width increase as the coherent regime is approached.}
  \label{fig:ratecross}
\end{figure}

By using the $\delta$-identity
\begin{equation}
    \delta \left(f(x)\right) = \sum_i\frac{\delta(x-x_i)}{|f'(x_i)|}\qquad \text{where}\qquad f(x_i) = 0,
\end{equation}
together with Eq.~\eqref{eq:Deltav}, Eq.~\eqref{eq:contp} becomes
\begin{equation}
    P(\Delta') = \frac{1}{m_a} f\left(\frac{\delta_k- \Delta'}{m_a}\right).
    \label{eq:Ptov}
\end{equation}
As a benchmark, we specialize to a one-dimensional Maxwell-Boltzmann distribution for the velocity $v$~\footnote{Strictly speaking, $v$ denotes $v_\parallel$, namely the projection of the axion velocity along the photon propagation direction. As discussed in Sec.~\ref{subsubsec:comment}, we neglect this distinction in the present one-dimensional treatment.}
\begin{equation}
    f(v) = \frac{1}{\sqrt{2\pi}\sigma_v}\exp\left(-\frac{v^2}{2\sigma_v^2}\right).
    \label{eq:mb}
\end{equation}
Then, the kernel in Eq.~\eqref{eq:Pdelta} can be evaluated analytically 
\begin{equation}
    K(\tau) = \kappa^2\int d\Delta' \frac{1}{\sqrt{2\pi}\sigma_\Delta}\exp\left(-\frac{(\Delta' - \delta_k)^2}{2\sigma_\Delta^2} + i\Delta'\tau\right) = \kappa^2 \exp\left( -\frac{\sigma_\Delta^2 \tau^2}{2} + i\delta_k\tau\right)
    \label{eq:kexp}
\end{equation}
where we have used Eq.~\eqref{eq:Ptov} and defined $\sigma_\Delta \equiv m_a \sigma_v$. The kernel is exponentially suppressed for $\tau \sim \sigma_\Delta^{-1}$, which indeed coincides with the photon crossing time. \\
\newline
When  $\alpha_k^\text{sym}(t)$ varies negligibly over the correlation time of the kernel, Eq.~\eqref{eq:mark} gives $\alpha_k^\text{sym}(t) \sim e^{s_k^\text{in}t}$, where
\begin{equation}
    s^\text{in}_k = \sqrt{\frac{\pi}{2}}\frac{\kappa^2}{\sigma_\Delta} \exp\left(-\frac{\delta_k^2}{2\sigma_\Delta^2}\right) \left[1 + i ~\text{erfi}\left(\frac{\delta_k}{\sqrt{2}\sigma_\Delta}\right)\right],
    \label{eq:skin}
\end{equation}
with $\text{erfi}$ the imaginary error function. Its real part determines the exponential growth rate
\begin{equation}
    \text{Re}\,[s_k^\text{in}] = \sqrt{\frac{\pi}{2}}\frac{\kappa^2}{\sigma_\Delta} \exp\left(-\frac{\delta_k^2}{2\sigma_\Delta^2}\right),
    \label{eq:gammakin}
\end{equation}
which perfectly agrees with the exponent in Eq.~\eqref{eq:mark2}, as 
\begin{equation}
    \pi \kappa^2  P(0)  = \sqrt\frac{\pi}{2} \frac{\kappa^2}{m_a\sigma_v} \exp\left(-\frac{\delta_k^2}{2\sigma_\Delta^2}\right) .
    \label{eq:Gammaincoh2}
\end{equation}
For $\delta_k=0$, this result also reproduces parametrically the qualitative estimate obtained in Eq.~\eqref{eq:schem}. More generally, Eq.~\eqref{eq:gammakin} shows that efficient amplification occurs only for photon momenta satisfying $|\delta_k| \lesssim \sigma_\Delta$. Equivalently, the axion velocity required to satisfy the resonance condition must lie within the appreciably populated region of the velocity distribution. For the Maxwell-Boltzmann distribution of Eq.~\eqref{eq:mb}, $f(v)$ is formally nonzero for every velocity, hence the growth rate is technically nonzero for every $k$. However, it is exponentially suppressed in the tails. A physical halo distribution also possesses an escape-velocity cutoff, in which case the resonance has finite support~\cite{ParticleDataGroup:2024cfk}.\\
\newline
The physical interpretation is transparent. When the photon amplitude changes only weakly during the crossing of a single axion coherence domain, the different momentum components dephase before substantial coherent amplification can develop. The photon then accumulates many small enhancements while traversing effectively incoherent regions. The cumulative effect remains exponential, but its exponent scales as $g_{a\gamma}^2$ rather than $g_{a\gamma}$.\\
\newline
This scaling is not specific to the Maxwell-Boltzmann distribution adopted above. Different choices of $f(v)$ may modify the detailed dependence of the growth rate on the momentum mismatch, alter the relevant correlation timescale, and change numerical prefactors. Nevertheless, in the incoherent regime, the overall dependence on $\kappa^2$ is generic. Parametrically, one indeed expects $\text{Re}\,[s_k^\text{in}] \sim \kappa^2\tau_\text{coh}$, where $\tau_\text{coh}$ denotes the correlation time of the kernel associated with the chosen velocity distribution. This provides a more general formulation of the heuristic argument in Eq.~\eqref{eq:schem} and supports the intuition that in the incoherent regime, we should recover a Boltzmann/kinetic-like behavior, governed by $\kappa^2 \propto g_{a\gamma}^2$.\\
\newline
The results obtained so far support the qualitative picture developed at the beginning of Sec.~\ref{sec:super}. However, the role of the parameter $\Theta$ introduced in Eq.~\eqref{eq:Theta} has not yet emerged explicitly from the calculation. This is a direct consequence of having assumed that $\alpha_k^\text{sym} (t)$ changes negligibly over the correlation time of the kernel. This effectively restricts the analysis to the regime in which dephasing occurs before significant coherent amplification can develop, corresponding parametrically to $\Theta\ll1$.\\
Equation~\eqref{eq:gammakin} nevertheless shows that the photon amplitude continues to grow exponentially even in this incoherent regime, albeit with an exponent that differs from the coherent result in Sec.~\ref{sec:coh}. To describe continuously the crossover between coherent and incoherent amplification, we therefore return to Eq.~\eqref{eq:volt} (setting again $S=0$), and look for its asymptotic exponential behavior, namely $\alpha_k^\text{sym }(t) \sim e^{s_kt}$.~\footnote{In the coherent regime, $s_k\tau_c^\text{cross}\gg1$. Therefore, the asymptotic exponential regime is reached on timescales much shorter than the photon crossing time of an axion coherence patch.}
Then, we obtain
\begin{equation}
    s_k = \int d\tau K(\tau)e^{-s_k\tau} = \sqrt{\frac{\pi}{2}}\frac{\kappa^2}{\sigma_\Delta}  \exp\left[\frac{(s_k - i \delta_k)^2}{2\sigma_\Delta^2}\right]\text{erfc}\left(\frac{s_k - i\delta_k}{\sqrt{2}\sigma_\Delta}\right),
    \label{eq:implicit}
\end{equation}
where $\text{erfc}(x)$ is the complementary error function. \\
Equation~\eqref{eq:implicit} is the central result of this work. It provides an implicit equation for the complex exponent $s_k$ and continuously connects the coherent regime discussed in Sec.~\ref{sec:coh} to the incoherent, kinetic-like regime derived above.\\
\newline
Notice that the quantity $s_k$ appearing in Eq.~\eqref{eq:implicit} denotes the growth exponent determined self-consistently by the implicit equation, and should not in general be identified with the coherent growth rate derived in Sec.~\ref{sec:coh}. It proves therefore convenient to actually deal with the ratio
\begin{equation}
    \tilde \Theta \equiv \frac{\kappa}{\sigma_\Delta} = 6.3\cdot 10^{-14} \left(\frac{g_{a\gamma}}{10^{-10}~\text{GeV}^{-1}}\right)\left(\frac{\rho_a}{\rho_\text{DM}}\right)^{1/2}\left(\frac{10^{-6}~\text{eV}}{m_a}\right)\left(\frac{10^{-3}}{\sigma_v}\right),
    \label{eq:tt}
\end{equation}
which corresponds to the value of $\Theta$ at the center of the resonance band in the Mathieu-like analysis of Sec.~\ref{sec:coh}.
For convenience, we also define the following dimensionless quantities
\begin{equation}
    x_k \equiv \frac{s_k}{\kappa}, \qquad d_k = \frac{\delta_k}{\sigma_\Delta},
    \label{eq:ss}
\end{equation}
which bring Eq.~\eqref{eq:implicit} to a dimensionless form:
\begin{equation}
    x_k = \sqrt{\frac{\pi}{2}} \tilde \Theta~\exp\left[\frac{(\tilde\Theta x_k - i d_k)^2}{2}\right]\text{erfc}\left(\frac{\tilde \Theta x_k - i d_k}{\sqrt{2}}\right).
    \label{eq:implicitadim}
\end{equation}
Equation~\eqref{eq:implicitadim} provides a continuous interpolation between the coherent and incoherent limits. We now analyze these two asymptotic regimes separately:
\begin{itemize}
    \item \textit{Coherent case}, $\tilde \Theta \gg 1$. Using the asymptotic expansion
    \begin{equation}
        \exp \left[\frac{Z^2}{2}\right]\text{erfc}\left(\frac{Z}{\sqrt{2}}\right) \sim \sqrt{\frac{2}{\pi}}\frac{1}{Z} + ... \qquad \text{for}~ |Z|\gg 1,
        \label{eq:expZ}
    \end{equation}
    Eq.~\eqref{eq:implicitadim} reduces to the simple equation
    \begin{equation}
        x_k \simeq \frac{1}{x_k - i \frac{d_k}{\tilde\Theta}}.
    \end{equation}
    The growing solution, corresponding to the branch with positive real part, is
    \begin{equation}
        x_k = \sqrt{1 - \left(\frac{d_k}{2\tilde\Theta}\right)^2} + i \frac{d_k}{2\tilde\Theta}.
        \label{eq:xc}
    \end{equation}
    Its real part exactly reproduces the coherent growing eigenvalue obtained in Eqs.~\eqref{eq:mut} and \eqref{eq:mutp}.~\footnote{The asymptotic expansion in Eq.~\eqref{eq:expZ} is valid provided that $|Z|\gg1$ and $|\text{arg}~Z|< 3\pi/4$. 
    One can readily verify \textit{a posteriori} that, when evaluated on the solution in Eq.~\eqref{eq:xc}, the argument entering Eq.~\eqref{eq:implicitadim} satisfies both conditions.} Thus, for $\tilde \Theta \gg 1$, Eq.~\eqref{eq:implicitadim} consistently recovers the coherent regime, and $s_k \propto g_{a\gamma}$.     
    \item \textit{Incoherent case}, $\tilde \Theta \ll 1$. Keeping $d_k$ fixed, Eq.~\eqref{eq:implicitadim} gives:
    \begin{equation}
        x_k \sim \sqrt{\frac{\pi}{2}}\tilde \Theta \exp\left[- \frac{d_k^2}{2}\right] \left[1 + i \,\text{erfi}\left(\frac{d_k}{\sqrt{2}}\right)\right] + \mathcal{O}(\tilde \Theta^3),
        \label{eq:inclim}
    \end{equation}
    which, upon the substitutions in Eqs.~\eqref{eq:tt} and \eqref{eq:ss} and the identity $\text{erfc}(-ix) = 1 + i\,\text{erfi}(x)$, reduces exactly to Eq.~\eqref{eq:skin}. In the incoherent regime, thus, $s_k \propto g_{a\gamma}^2$, in agreement with the kinetic interpretation developed above.
\end{itemize}
Equation~\eqref{eq:implicitadim} provides a smooth interpolation between the coherent and incoherent regimes. As shown in Appendix~\ref{app:full}, the simplified treatment adopted here, based on the ansatz in Eq.~\eqref{eq:A+} and on Eqs.~\eqref{eq:alphatot}-\eqref{eq:betasj}, is independently justified in both asymptotic limits, so that the corresponding coherent and incoherent results remain robust. In the crossover region, however, the presence of additional mode couplings (see Eqs.~\eqref{eq:afu}-\eqref{eq:bfu}) can introduce quantitative corrections. Equation~\eqref{eq:implicitadim} should therefore be regarded as an approximate description of the interpolation between the two correctly recovered asymptotic regimes. \\
\newline
We conclude this section with a series of figures. In Fig.~\ref{fig:theta_regions}, we plot the curve $\tilde \Theta =1$, with $\tilde \Theta$ defined in Eq.~\eqref{eq:tt}, for several values of the axion density $\rho_a$, while fixing the velocity dispersion to the benchmark value adopted in Eq.~\eqref{eq:tt}. Broadly speaking, this contour separates the $(g_{a\gamma},m_a)$ plane into a coherent region, $\tilde \Theta >1$, and an incoherent region, $\tilde \Theta \leq1$. For simplicity, the intermediate crossover described by Eq.~\eqref{eq:implicitadim} is not explicitly displayed. For reference, we also show the conventional QCD axion band, together with the KSVZ~\cite{Kim:1979, SHIFMAN1980493} and DFSZ~\cite{DINE1981199, Zhitnitsky:1980tq} benchmark models, obtained from the ``AxionLimits" GitHub repository.~\cite{AxionLimits}.\\
\newline
For the density and velocity dispersion characteristic of ordinary virialized Galactic DM, the axion parameter space lies extremely deep in the incoherent regime.
Larger values of $\tilde\Theta$ may be obtained in overdense axion environments, such as axion minihalos and miniclusters~\cite{Xiao_2021, Eggemeier_2020}, axion stars~\cite{Braaten_2016,Visinelli_2018} or DM spikes around black holes~\cite{Gondolo_1999, Ullio_2001, Bertone_2005, Shen:2023kkm, Lacroix:2018zmg}, which we study in a companion work~\cite{Smarra:2026}. Nevertheless, for the representative densities shown in Fig.~\ref{fig:theta_regions}, a large fraction of the phenomenologically relevant parameter space remains in the incoherent regime. We stress that some of these configurations, in particular axion stars and other highly coherent bound objects, need not be described by the same virialized velocity distribution assumed in Eq.~\eqref{eq:mb}. Therefore, the comparison with Fig.~\ref{fig:theta_regions} should be regarded as purely indicative.\\
\newline
It is instructive to estimate the density required for a standard QCD axion to enter the coherent regime. Parametrically, the axion-photon coupling satisfies~\cite{Di_Luzio_2020}
\begin{equation}
    g_{a\gamma}^\text{QCD} \sim \frac{\alpha}{2\pi} \frac{m_a}{\Lambda^2_\text{QCD}} C_{a\gamma},
\end{equation}
where $C_{a\gamma}$ is a model-dependent order one coefficient and $\Lambda_\text{QCD} \sim 0.1~\text{GeV}$ is the QCD confining scale. Equating this expression to the coupling corresponding to $\tilde\Theta=1$, namely $g_{a\gamma}^{\tilde \Theta = 1}$, obtained by inverting Eq.~\eqref{eq:tt}, gives
\begin{equation}
    \rho_a \sim 7.8 \cdot 10^{37}~\text{GeV}/\text{cm}^3 \frac{1}{C_{a\gamma}^2}\left(\frac{\sigma_v}{10^{-3}}\right)^2\left(\frac{\Lambda_\text{QCD}}{0.1~\text{GeV}}\right)^4.
    \label{eq:densQCD}
\end{equation}
The axion mass cancels from this estimate because both $g_{a\gamma}^\text{QCD}$ and $g_{a\gamma}^{\tilde \Theta = 1}$ scale linearly with $m_a$.
Equation~\eqref{eq:densQCD} therefore illustrates how extreme the densities required to reach the coherent regime are for a QCD axion in a virialized environment.\\
Among the most extreme axion configurations are superradiant boson clouds around spinning black holes~\cite{Brito:2017short, Brito:2017long, Brito:2020, Ellis:2026gwt}.
These systems can develop very large occupation numbers and substantial local densities, although their interpretation within Fig.~\ref{fig:theta_regions} requires some care, as they should not be interpreted directly in terms of the virialized Maxwell-Boltzmann distribution assumed in the estimates above. ~\footnote{A simple order-of-magnitude estimate illustrates the characteristic densities involved. Considering a benchmark value for the gravitational fine-structure constant $\alpha \sim 0.1$ and a black hole mass of $ M_\text{BH} =10~\text{M}_\odot$, the boson cloud mass is approximately $M_\text{cloud} \sim \alpha M_\text{BH}$, with typical cloud radius of $R_\text{cloud}\sim M_\text{BH}/\alpha^2M_\text{pl}^2$, where $M_\text{pl} = 1.22\cdot 10^{19}~\text{GeV}$ is the Planck mass. Putting all together, we find
\begin{equation}
    \rho_\text{cloud} \sim \frac{M_\text{cloud}}{R_\text{cloud}^3} \sim 3.4\cdot 10^{32}~\text{GeV}/\text{cm}^3 \left(\frac{10~\text{M}_\odot}{M_\text{BH}}\right)^2.
\end{equation} 
The corresponding boson mass is $m_a \sim \alpha M_\text{pl}^2/M_\text{BH} \sim 1.3\cdot 10^{-12}~\text{eV}$. Although this density is still smaller than the value in Eq.~\eqref{eq:densQCD}, we remark that superradiant clouds should not be interpreted as ordinary virialized Maxwell-Boltzmann distributions. Indeed, the cloud occupies one or a small number of coherent bound states, so its phase-space structure differs qualitatively from the picture developed in Sec.~\ref{sec:cont}. A dedicated treatment is therefore required to determine precisely how the coherent-incoherent criterion developed here applies to superradiant clouds. }\\
\newline
As a remark, the regions displayed in Fig.~\ref{fig:theta_regions} should not be interpreted as universally accessible to electromagnetic amplification. Depending on the astrophysical environment, plasma effects can kinematically suppress axion decay into propagating photons when the relevant photon plasma frequency is too large. The parameter space shown in Fig.~\ref{fig:theta_regions} is therefore intended only to illustrate the coherent--incoherent transition and must be supplemented by the appropriate environmental conditions for any specific application; see also Footnote~\ref{foot:plasma}.
\\
\newline
Figure~\ref{fig:growthrate} shows the solution $x_k$ of the implicit equation~\eqref{eq:implicitadim} as a function of $\tilde \Theta$. For simplicity, we restrict to exact resonance, $d_k=0$. The solid curve, corresponding to the full solution of Eq.~\eqref{eq:implicitadim}, smoothly interpolates between the incoherent and coherent limits derived in Eqs.~\eqref{eq:inclim} and \eqref{eq:xc}, respectively.
\\
\newline
Finally, Fig.~\ref{fig:ratecross} shows $\text{Re}\,[x_k]$ as a function of the normalized mismatch $d_k$ for selected values of $\tilde \Theta$. As $\tilde \Theta$ increases, the resonance profile changes smoothly from the Gaussian form characteristic of the incoherent regime to the square-root behavior of the coherent limit (see Eqs.~\eqref{eq:inclim} and \eqref{eq:xc}). At the same time, the peak growth rate increases, consistently with the behavior displayed in Fig.~\ref{fig:growthrate}. \\
\newline
Notice also that the profile of $\text{Re}\,[x_k]$ broadens as $\tilde \Theta$ increases. This behavior can be understood by comparing the amplification and dephasing timescales. In the incoherent regime, amplification is slow enough for the photon to resolve the momentum differences among the axion modes, and the range of efficiently amplified photon momenta is therefore controlled by the width $\sigma_\Delta$ of the axion momentum distribution. In the coherent regime, instead, amplification occurs on a timescale $\sim\kappa^{-1}$, shorter than the time required for nearby momentum modes to appreciably dephase. Momentum mismatches satisfying $|\delta_k|\lesssim2\kappa$ therefore cannot suppress the growth before significant amplification develops. In this sense, an increasingly broad range of axion momenta becomes dynamically indistinguishable on the amplification timescale as $\tilde\Theta$ increases. Equivalently, in terms of the normalized mismatch $d_k$, the coherent resonance extends over $|d_k|\lesssim2\tilde\Theta$, corresponding to the finite width of the first Mathieu instability band (see Sec.~\ref{sec:coh}).

\section{Conclusions and outlook}\label{sec:concl}

In this work, we have studied stimulated axion decay into photons directly from the classical Maxwell equations, with particular emphasis on the finite coherence of the axion field. For a coherent axion background, the standard parametric resonance treatment predicts an exponentially growing photon population, with growth exponent linear in the axion-photon coupling, $s\propto g_{a\gamma}$. By contrast, a kinetic description, for instance based on Boltzmann equations, naturally involves the squared matrix element of the relevant interaction process, and therefore would point to a characteristic rate scaling as $s\propto g_{a\gamma}^2$ (see also our companion work~\cite{Smarra:2026}). The main goal of this work has been to understand the origin of these seemingly different scalings and to clarify how they are related. We have shown that they arise as distinct limits of the same underlying dynamics, determined by the coherence properties of the axion field.\\
\newline
In Sec.~\ref{sec:coh}, we reviewed the standard case of a coherent, zero-momentum axion background. In this limit, the modified Maxwell equations reduce to a Mathieu equation, and the photon field develops an exponential instability with growth exponent $s\propto g_{a\gamma}$. We then showed that the presence of a finite axion momentum does not by itself suppress the instability, along the lines of Ref.~\cite{Arza_2020}.\\
The situation changes when the axion field contains a distribution of momentum modes. In Sec.~\ref{sec:super}, we considered a discrete superposition of axion modes and showed that different axion components couple the signal photon to different idler modes, which accumulate different phases during the evolution. The resulting evolution of the signal photon can be expressed in terms of a \textit{kernel} encoding the axion momentum distribution. The decay of this kernel is controlled by the dephasing of the different momentum components and occurs over a characteristic timescale of order the photon crossing time of an axion coherence domain. \\
\newline
This observation provides a simple physical interpretation of the coherent-to-incoherent transition. If the amplification timescale is shorter than the time required for a photon to traverse an axion coherence domain, significant growth occurs before the different momentum components have time to dephase. The photon therefore probes an approximately coherent axion background, and the standard parametric-resonance behavior $s\propto g_{a\gamma}$ is recovered. Conversely, if the amplification is slower than the dephasing induced by the axion momentum spread, the photon effectively samples several incoherent components before appreciable growth takes place. The coherent buildup is then suppressed and an effectively kinetic regime emerges, characterized by $s \propto g_{a\gamma}^2$. \\
\newline
In Sec.~\ref{sec:cont}, we made this transition quantitative by considering a continuous axion velocity distribution. For a Maxwell-Boltzmann distribution, the dephasing kernel can be evaluated analytically, allowing us to derive an implicit equation for the photon growth exponent that interpolates between the coherent and incoherent regimes. The transition is controlled by the dimensionless parameter $\tilde \Theta = \kappa/\sigma_\Delta$, where $\kappa$ is the coherent on-resonance growth rate and $\sigma_\Delta$ characterizes the width of the axion momentum dispersion. Taking the limits $\tilde \Theta \gg 1$ and $\tilde \Theta\ll1$, we recovered the coherent and incoherent regimes, respectively. Although the detailed shape of the crossover depends on the axion momentum distribution, we argued that its qualitative structure, and in particular the dependence on $g_{a\gamma}$ in the two asymptotic limits, is largely insensitive to the specific distribution adopted. \\
\newline
Our results therefore provide a direct classical-field interpretation of the transition between the parametric resonance and the kinetic regimes. In particular, the $g_{a\gamma}^2$ scaling emerges naturally from the Maxwell equations once the finite momentum dispersion and the associated dephasing are taken into account.
Conversely, the linear dependence on $g_{a\gamma}$ is recovered whenever the instability develops rapidly enough that the axion background remains effectively coherent over the relevant amplification scale. The coherent and kinetic descriptions should therefore be regarded as two limiting regimes of the same physical process.\\
\newline
Several directions remain open for extending the present analysis. First, throughout this work, we have treated the axion background as fixed and neglected the backreaction of the amplified electromagnetic field. This approximation will eventually break down once a significant fraction of the axion energy is transferred to photons. Incorporating backreaction therefore constitutes a natural direction for future work. It would also be interesting to extend the present one-dimensional treatment to a fully three-dimensional system. \\
An important further direction is to complement the analytical results presented here with numerical simulations. As discussed in Sec.~\ref{sec:cont} and in more detail in Appendix~\ref{app:full}, the implicit equation for the photon growth exponent derived in Eq.~\eqref{eq:implicitadim} correctly reproduces the coherent and incoherent asymptotic limits. In the intermediate regime, neglected mode couplings may introduce quantitative corrections to the crossover shown in Fig.~\ref{fig:growthrate}. A numerical analysis of the full coupled system in Eqs.~\eqref{eq:afu}-\eqref{eq:bfu} would be needed to determine the detailed shape of this transition, which we leave for future work. 
Such an analysis would also provide a natural framework for exploring different axion momentum distributions and assessing their impact on the coherent-to-incoherent transition.
\\
Finally, from a more formal perspective, it would be valuable to recover the same crossover within a non-equilibrium quantum field theory framework. In particular, deriving the coherent and incoherent limits starting from the Kadanoff-Baym equations~\cite{Berges_2004} would provide a direct connection between the classical treatment developed here and a microscopic quantum description of the system. We plan to investigate this question in future work.

\section*{Acknowledgements}
CS thanks Mathias Becker, Luca Di Luzio and J\"org J\"ackel for insightful discussions and valuable feedback on the manuscript. He also thanks J\"urgen Berges for discussions on the non-equilibrium QFT framework, and Philip S\o rensen for stimulating discussions.
CS acknowledges the Leinweber Institute for Theoretical Physics of UC Berkeley and the Lawrence Berkeley National Laboratory for hospitality during the initial development of this work.
The work of CS was supported by the European Union – NextGeneration EU, mission 4, component 1, CUP C93C24004930006.  This project has received funding from the European Union's Horizon Europe research and innovation programme under the Marie Skłodowska-Curie Staff Exchange  grant agreement No 101086085 – ASYMMETRY. ET is supported by the European Union’s Horizon 2020
research and innovation programme under the Marie Skłodowska-Curie grant agreement No 101204903 (APARAX).
CS acknowledges the use of ChatGPT to check derivations, rephrase text and generate figure-related code.

\newpage
\appendix
\section{Coherence length, coherence time and local density}\label{app:local}
In Sec.~\ref{sec:coh}, we argued that Eq.~\eqref{eq:at} provides an appropriate description of a virialized axion field when restricting to lengthscales smaller than the axion coherence length, or equivalently to timescales shorter than the axion coherence time. Later, in Eq.~\eqref{eq:kj} of Sec.~\ref{sec:discrmom}, we introduced $\rho_j$ as the \textit{averaged} energy density carried by the axion momentum mode $p_j$.  In this Appendix, we clarify the previous statements.\\
\newline
First, let us consider a generic axion field, written for simplicity as a discrete superposition of momentum modes, as in Eq.~\eqref{eq:axcomp}. As we are dealing with non-relativistic axions, it is convenient to factor out the rapid oscillations at the axion mass scale: 
\begin{equation}
    a (z, t) = \frac{1}{2}e^{-im_at }\sum_j a_j e^{-i\delta\omega_j t + ip_j z - i\phi_j} + \text{c.c.} ,
    \label{eq:pt}
\end{equation}
where $\delta\omega_j = \omega_j - m_a$. 
By introducing the slowly varying complex envelope 
\begin{equation}
    \Psi(z,t) \equiv \sum_j a_j e^{-i\delta\omega_j t + ip_j z - i\phi_j} \qquad \vartheta(z, t ) \equiv \text{arg}~\Psi(z, t),
    \label{eq:psiz}
\end{equation}
Eq.~\eqref{eq:pt} can be written as
\begin{equation}
    a(z, t) = |\Psi(z, t)| \cos(m_a t - \vartheta(z, t)).
    \label{eq:acos}
\end{equation}
The quantity $\Psi(z, t)$ encodes the slow modulation of the axion field, and naturally determines its coherence properties. Indeed, for
\begin{equation}
    |\Delta z|\lesssim  L_c,\qquad  |\Delta t| \lesssim \tau_c ,
\end{equation}
with $L_c$ and $\tau_c$ defined in Eqs.~\eqref{eq:Lc} and \eqref{eq:tc}, the relative phases of the different momentum modes vary only weakly. Within such a coherence patch, $|\Psi|$ and $\vartheta$ may therefore be regarded approximately as constant, and the axion field is well approximated by a single cosine with fixed amplitude and phase. This motivates the discussion below Eq.~\eqref{eq:at}.\\
Instead, on scales larger than $\tau_c$ or $L_c$, the different momentum modes in Eq.~\eqref{eq:psiz} progressively dephase. In that regime, the spatial and temporal dependence of the envelope $\Psi(z,t)$ can no longer be neglected, and the full expression in Eq.~\eqref{eq:acos} must be retained.\\
\newline
The axion energy density follows from the stress energy tensor:
\begin{equation}
    \rho_a = T^{00} = \frac{1}{2}\left(\partial_0 a\right)^2 + \frac{1}{2}\left(\nabla a\right)^2 + \frac{1}{2}m_a^2 a^2.
\end{equation}
For non-relativistic axions, the gradient contribution is suppressed by powers of the velocity and may be neglected at leading order. Moreover, since the envelope varies on timescales much longer than $m_a^{-1}$,
\begin{equation}
    \rho_a \simeq \frac{1}{2}m_a^2|\Psi(z, t)|^2.
    \label{eq:rhopsi}
\end{equation}
For a single momentum mode, as considered in Sec.~\ref{sec:coh}, this immediately reproduces the normalization of the axion amplitude used in Eq.~\eqref{eq:at}. However, when a superposition of modes is considered, the situation is more subtle. Indeed, in that case,
\begin{equation}
    |\Psi(z, t)|^2 = \sum_j a_j^2 + 2 \sum_{j>l} a_j a_l \cos(\Phi_j(z,t) - \Phi_l(z,t)),
    \label{eq:int}
\end{equation}
where we have defined for simplicity 
\begin{equation}
    \Phi_j (z, t) = \delta \omega_j t - p_j z + \phi_j.
\end{equation}
Thus, the local axion density is determined not only by the sum of individual momentum components, but also by their mutual interference. Different coherence patches therefore generally correspond to different local values of $\rho_a$.~\footnote{For detailed discussions of these interference-induced density fluctuations, see Refs.~\cite{Foster:2018,Centers:2021}. A useful visualization of the resulting coherence-patch structure in the fuzzy-DM regime can be found in Ref.~\cite{Schive:2014}. Such spatial and temporal coherence properties are also relevant for Pulsar Timing Array searches for ultralight DM; see, e.g., Refs.~\cite{Porayko:2018,Castillo:2022,Smarra:2023egv,Foster:2026mvs}.} Upon averaging over many realizations, or equivalently over sufficiently large spacetime regions, the interference terms vanish. We then obtain:
\begin{equation}
    \langle \rho_a\rangle = \frac{1}{2}m_a^2 \sum_j a_j^2 = \sum_j\rho_j,
    \label{eq:avrhoa}
\end{equation}
where we naturally interpret $\rho_j$ as the \textit{averaged} energy density associated with the axion momentum mode $p_j$, as done in Eq.~\eqref{eq:kj}. Without this averaging, the sum $\sum_j\rho_j$ would not coincide with the local axion density, since the latter also contains the interference terms in Eq.~\eqref{eq:int}.\\
\newline
The distinction between the local and averaged densities becomes particularly relevant when comparing the coherent and incoherent regimes discussed in the main text.
As discussed more explicitly in Appendix~\ref{app:full}, in the incoherent regime, the relative phases of the different axion momentum components dephase on timescales short compared with the amplification timescale, and only the diagonal terms proportional to $a_j^2$ survive. The quantities $\rho_j$ entering the corresponding description should therefore be interpreted as the averaged energy densities carried by the individual momentum modes.
Conversely, in the coherent treatment in Sec.~\ref{sec:coh}, one effectively focuses on a single coherence patch. The density $\rho_a$ entering the growth rate should therefore be understood as the local axion density of that patch, including the contribution from interference among the underlying momentum modes. This simply rescales the local value of the coherent growth exponent $\kappa$ in Eq.~\eqref{eq:pars0} accordingly.

\section{Full signal-idler equations}\label{app:full}
In Sec.~\ref{sec:discrmom}, we solved the modified Maxwell equations using the ansatz in Eq.~\eqref{eq:A+}, which contains a single signal mode propagating along the positive $z$-direction with momentum $k$, together with a set of idler modes, with momenta $q_j = p_j - k$. As already discussed, this ansatz is not closed under the Maxwell evolution. Indeed, an idler photon with momentum $q_j$, generated by interaction with the axion mode $p_j$, can subsequently interact with a different axion component $p_l$, thereby sourcing a new signal mode with momentum
\begin{equation}
    k_{lj} = p_l - q_j = k + p_l - p_j.
\end{equation}
Therefore, in the full system, $\beta_j$ does not couple exclusively to the original signal mode $\alpha_k$, as assumed in Eqs.~\eqref{eq:alphatot}-\eqref{eq:betasj}, but also to the additional photon modes $\alpha_{k+ p_l - p_j}$. Therefore, the appropriate ansatz is
\begin{equation}
    \vec{A}_\lambda^+ = \sum_l\vec{\epsilon}_\lambda \alpha_{r_l}(t) e^{ir_lz}e^{-ir_lt} + \sum_j\vec\epsilon_{-\lambda}\beta_{q_j}(t) e^{iq_jz} e^{-i|q_j|t},
    \label{eq:Acorr}
\end{equation}
where we have introduced a different notation from Eq.~\eqref{eq:A+} to distinguish the general system considered here.\\
\newline
Substituting Eq.~\eqref{eq:Acorr} into Eq.~\eqref{eq:Amom} and following the same steps as in Sec.~\ref{sec:discrmom}, we obtain:
\begin{align}
        &\dot\alpha_k =  \sum_j\kappa_{\alpha j}  \beta_{p_j-k}^* e^{-i\phi_j}e^{i\Delta_j  (k) t}  \label{eq:afu}\\
        &\dot\beta^*_{q_j} =\sum_l\kappa_{\beta l}(q_j)\alpha_{p_l - q_j} e^{i\phi_l}e^{-i\Delta_{l}(p_l - q_j) t},
        \label{eq:bfu}
\end{align}
where we have chosen to follow the evolution of the signal mode with momentum $r_l = k$ and defined 
\begin{equation}
    \kappa_{\alpha j}(k) \equiv \lambda\frac{g_{a\gamma}a_j}{4}\frac{|p_j - k|}{k}(\omega_j + p_j) \qquad \kappa_{\beta l} (q_j) \equiv \lambda\frac{g_{a\gamma}a_l}{4}\frac{p_l - q_j}{|q_j|}(\omega_l - p_l),
\end{equation}
while the detunings read
\begin{equation}
    \Delta_j(x) = x + |p_j-x| - \omega_j.
    \label{eq:gend}
\end{equation}
Unlike the simplified system in Eqs.~\eqref{eq:alphatot}-\eqref{eq:betasj}, now each idler mode $\beta_{q_j}$ couples to several $\alpha_{r_l}$ modes. As in Sec.~\ref{sec:discrmom}, we may formally integrate Eq.~\eqref{eq:bfu} and plug it back into Eq.~\eqref{eq:afu}. This yields
\begin{equation}
    \dot\alpha_k(t) = \sum_{j,l}\kappa_j \kappa_l e^{i(\phi_l - \phi_j)}e^{i\left[\Delta_j(k)- \Delta_l(k + p_l - p_j)   \right] t } \int_0^t d\tau ~\alpha_{k + p_l - p_j}(t-\tau) e^{i\Delta_l(k + p_l - p_j)\tau},
    \label{eq:voltg}
\end{equation}
where we have approximated 
\begin{equation}
    \kappa_{\alpha j}(k) \sim \lambda \frac{g_{a\gamma} a_j m_a}{4} =\lambda \kappa_j \qquad \kappa_{\beta l}(q_j) \sim \lambda \frac{g_{a\gamma} a_l m_a}{4} =\lambda \kappa_l ,
\end{equation}
with $\kappa_j$ as defined in Eq.~\eqref{eq:kj}. Again, we have conveniently neglected the contribution from the initial occupation of the idler modes (see Eq.~\eqref{eq:volt}). \\
\newline
Equation~\eqref{eq:voltg} is the generalization of Eq.~\eqref{eq:volt}. The latter is recovered by retaining only the diagonal terms $j = l$, whereas in the full problem the off-diagonal contributions with $j\neq l$ are also present. It is therefore important to understand under which conditions the simplified treatment  in Secs.~\ref{sec:super}-\ref{sec:cont} remains justified. \\
\newline
In particular, let us write explicitly the difference between the detunings appearing in Eq.~\eqref{eq:voltg}:
\begin{equation}
    \Omega_{jl} \equiv \Delta_j(k) - \Delta_l( k + p_l - p_j) = (\omega_l - p_l) - (\omega_j - p_j),
\end{equation}
where Eq.~\eqref{eq:gend} has been used in the last equality. The diagonal contributions, $j=l$, are precisely those retained in Eq.~\eqref{eq:volt}, and satisfy $\Omega_{jl} = 0$. By contrast, for the off-diagonal terms, $j \neq l$, the frequency difference is nonvanishing. For typical mode pairs, at leading order in the axion momentum, it can be estimated as: 
\begin{equation}
    |\Omega_{jl}| \sim |p_j - p_l| \sim \sigma_\Delta,
\end{equation}
where $\sigma_\Delta$, introduced in Eq.~\eqref{eq:kexp}, denotes the characteristic width of the axion momentum distribution. \\
\newline
Let us first consider the \textit{incoherent} regime, characterized by $\tilde \Theta \ll 1$. In this limit, the photon crosses many uncorrelated patches before being appreciably enhanced. The relevant evolution therefore takes place over times satisfying
\begin{equation}
    t \gg \tau_c^\text{cross} \Leftrightarrow t\gg\frac{1}{|\Omega_{jl}|}.
\end{equation}
Consequently, the rapidly oscillating factors in Eq.~\eqref{eq:voltg} suppress the off-diagonal contributions, and Eq.~\eqref{eq:volt} therefore provides a good description of the exponential enhancement in the \textit{incoherent} limit. \\
\newline
The situation is different in the \textit{coherent} regime, i.e. $\tilde \Theta \gg 1$. In this case, significant amplification occurs on timescales shorter than the photon crossing time of a coherence domain $\tau_c^\text{cross}$. Thus, in this regime, the off-diagonal terms do not average out and must, in principle, be retained. When $\tilde \Theta \gg 1$, the typical timescales we deal with are smaller than the photon crossing time:
\begin{equation}
    t \ll \tau_c^\text{cross} \Leftrightarrow \Omega_{jl}~t\ll1.
\end{equation}
Thus, we may approximate
\begin{equation}
    e^{i\Omega_{jl} t }\sim 1
\end{equation}
in Eq.~\eqref{eq:voltg}. For the same reason, we can further simplify
\begin{equation}
    e^{i\Delta_l(k + p_l - p_j)\tau} \sim e^{i\delta_k\tau} e^{i(p_l - 2p_j)\tau} \sim e^{i\delta_k\tau},
\end{equation}
thereby reducing Eq.~\eqref{eq:voltg}  to
\begin{equation}
    \dot\alpha_k(t) = \sum_{j,l}\kappa_j \kappa_l e^{i(\phi_l - \phi_j)} \int_0^t d\tau ~\alpha_{k + p_l - p_j}(t-\tau) e^{i\delta_k\tau}.
    \label{eq:voltgsimp}
\end{equation}
Differentiating with respect to time and using the Leibniz rule gives 
\begin{equation}
    \begin{split}
        &\ddot\alpha_k(t) = \sum_{j,l}\kappa_j \kappa_l e^{i(\phi_l - \phi_j)}\left[\alpha_{k + p_l - p_j}(t) + i\delta_k \int_0^t d\tau ~\alpha_{k + p_l - p_j}(t-\tau) e^{i\delta_k\tau}\right]\\
        &\ddot\alpha_k(t)  = \sum_{j,l}\kappa_j \kappa_l e^{i(\phi_l - \phi_j)}\alpha_{k + p_l - p_j}(t) + i\delta_k \dot \alpha_k(t),
    \end{split}
    \label{eq:so}
\end{equation}
where Eq.~\eqref{eq:voltgsimp} has been used in the second line. To solve this system, we introduce
\begin{equation}
    \mathcal{A}(\xi, t) = \sum_r \alpha_r(t) e^{ir\xi} \qquad \alpha_r(t) = \frac{1}{L}\int_0^Ld\xi ~\mathcal{A}(\xi, t) e^{-ir\xi},
\end{equation}
where $L$ is the size of the periodic box used to discretize the momenta. The quantity $\mathcal{A}(\xi, t)$ is simply the Fourier series of the right-moving photon field appearing in Eq.~\eqref{eq:Acorr}, with $\xi = z-t$. Then, Eq.~\eqref{eq:so} becomes
\begin{equation}
    \ddot{\mathcal{A}}(\xi,t)  = \kappa_\text{loc}^2~\mathcal{A}(\xi, t) + \left(2\partial_\xi - i m_a \right)\dot{\mathcal{A}}(\xi,t),
    \label{eq:ft}
\end{equation}
where the dots denote time derivatives at fixed $\xi$ and
\begin{equation}
    \kappa_\text{loc} = \left|\sum_{j}\kappa_j e^{i(p_j \xi - \phi_j)}\right|.
    \label{eq:Gammaxi}
\end{equation}
In general, the full signal-idler system in Eq.~\eqref{eq:so} couples $\alpha_k$ with modes $\alpha_{k + p_l - p_j}$. Therefore, the photon modes participating in the evolution are distributed over a momentum interval of $\sim m_a\sigma_v$ around $k$. Thus, we typically expect 
\begin{equation}
    \partial_\xi \dot {\mathcal{A}}(\xi, t) \sim i k \dot {\mathcal{A}}(\xi, t) + \mathcal{O}(m_a \sigma_v) \dot {\mathcal{A}}(\xi, t).
\end{equation}
Plugging this back into Eq.~\eqref{eq:ft}, we find
\begin{equation}
    \ddot{\mathcal{A}}(\xi,t)  = \kappa_\text{loc}^2\mathcal{A}(\xi, t) + i\delta_k \dot{\mathcal{A}}(\xi,t) + \mathcal{O}(m_a \sigma_v) \dot {\mathcal{A}}(\xi, t).
    \label{eq:ddA}
\end{equation}
Neglecting for the moment the last term and looking for solutions of the form $\mathcal{A}(\xi, t)\sim e^{s(\xi) t}$, we obtain
\begin{equation}
    s(\xi)^2 = \kappa_\text{loc}^2 + i\delta_k s(\xi),
    \label{eq:seq}
\end{equation}
whose growing root is
\begin{equation}
    s(\xi) = \sqrt{\kappa_\text{loc}^2 - \frac{\delta_k^2}{4}} + i\frac{\delta_k}{2}.
    \label{eq:sxi}
\end{equation}  
For simplicity, let us focus on sufficiently short times such that $\xi\simeq z$.~\footnote{This is also the regime of validity of the present approximation, which requires $t\ll\tau_c^\text{cross}$.} We then have
\begin{equation}
    \kappa_\text{loc}^2=   \left(\frac{g_{a\gamma} m_a}{4}\right)^2|\Psi(z, 0)|^2 = \frac{g_{a\gamma}^2 \rho_a^\text{loc}}{8},
    \label{eq:sumi}
\end{equation}
where we have used Eq.~\eqref{eq:int} and Eq.~\eqref{eq:rhopsi} to recast $|\Psi(z, t)|^2$ in terms of the axion density, which we denote by $\rho_a^\text{loc}$ to emphasize that it is the local, patch-dependent density rather than its ensemble-averaged value. \\
\newline
Equation~\eqref{eq:sxi}, together with Eq.~\eqref{eq:sumi}, has exactly the same structure as the coherent growth exponents derived in Eqs.~\eqref{eq:mutp} and ~\eqref{eq:xc}. Thus, in the coherent regime, the interference between different momentum modes cannot be neglected. Nevertheless, its net effect is simply to replace the averaged axion energy density by the local density of the coherence patch under consideration. Consequently, once $\kappa^2$ is replaced by $\kappa^2_\text{loc}$, the simplified signal-idler system of Eqs.~\eqref{eq:alphatot}-\eqref{eq:betasj} reproduces the result of the full system analyzed here in the deep coherent limit.\\
\newline
Furthermore, let us justify neglecting the $\mathcal O(m_a\sigma_v)\dot{\mathcal A}$ term in Eq.~\eqref{eq:ddA}. Indeed, this correction is potentially relevant close to resonance, where $|\delta_k| \lesssim m_a\sigma_v$. In that case, the structure of Eq.~\eqref{eq:ddA} becomes
\begin{equation}
    \ddot{\mathcal{A}}(\xi,t) \sim \kappa_\text{loc}^2 \mathcal{A}(\xi,t) + \mathcal{O}(m_a\sigma_v) \dot{\mathcal{A}}(\xi,t).
\end{equation}
Since the characteristic growth timescale is $\sim \kappa_\text{loc}^{-1}$, the second term is negligible when $\kappa_\text{loc} \gg m_a\sigma_v$, precisely the condition defining the coherent regime.\\
\newline
Therefore, the simplified treatment of Secs.~\ref{sec:super}-\ref{sec:cont} is justified in the two asymptotic regimes.
In the deep incoherent limit, the off-diagonal contributions dephase over timescales short compared with the amplification time and therefore average out dynamically. In the deep coherent limit, instead, they remain coherent but can be absorbed into the patch-dependent axion density, so that the simplified treatment remains valid after replacing the growth exponent $\kappa$ computed with the average density by its local counterpart $\kappa_\text{loc}$. \\
The intermediate crossover between these two limits is more subtle. In this regime, the off-diagonal contributions are neither fully coherent nor completely averaged out, and the detailed evolution can therefore depend on their interference. A complete characterization of individual realizations across the transition requires solving the full signal-idler system in  Eqs.~\eqref{eq:afu}-\eqref{eq:bfu}. We leave such a numerical investigation for future work.

\bibliography{biblio}

\end{document}